\documentclass[journal=langd5,manuscript=article]{achemso}
\DeclareSymbolFont{matha}{OML}{txmi}{m}{it}
\DeclareMathSymbol{\varv}{\mathord}{matha}{118}
\usepackage[version=3]{mhchem} 
\usepackage{amssymb}
\usepackage{xcolor}
\usepackage{soul,ulem}

\newcommand{\blu}[1]{{\color{black}{#1}}}

\author{Russell Kajouri}
\affiliation[ifpan]
{Institute of Physics, Polish Academy of Sciences, Al. Lotnik\'ow 32/46, 02-668 Warsaw, Poland}
\author{Panagiotis E. Theodorakis}
\affiliation[ifpan]
{Institute of Physics, Polish Academy of Sciences, Al. Lotnik\'ow 32/46, 02-668 Warsaw, Poland}
\email{panos@ifpan.edu.pl}
\author{Piotr Deuar}
\affiliation[ifpan]
{Institute of Physics, Polish Academy of Sciences, Al. Lotnik\'ow 32/46, 02-668 Warsaw, Poland}
\author{Rachid Bennacer}
\affiliation[ENS]{Université Paris-Saclay, ENS Paris-Saclay, CNRS, LMPS, \blu{4 Av. des Sciences, 91190 }Gif-sur-Yvette, France}
\author{Jan \v{Z}idek}
\affiliation{Central European Institute of Technology, Brno University of Technology, Purky\v{n}ova 656/123, 612 00 Brno, Czech Republic}
\author{Sergei A. Egorov}
\affiliation[University Virginia]
{Department of Chemistry, University of Virginia, Charlottesville, VA 22901, USA}
\alsoaffiliation[JoGu University]
{Institut f\"ur Physik, Johannes Gutenberg Universit\"at Mainz, 55099 Mainz, Germany}
\alsoaffiliation[Leibniz-Institut fuer polymerforschung]
{Leibniz-Institut f\"ur Polymerforschung, Institut Theorie der Polymere, Hohe Str. 6, 01069 Dresden, Germany}
\author{Andrey Milchev}
\affiliation{Bulgarian Academy of Sciences, Institute of Physical Chemistry, 1113 Sofia, 
Bulgaria}

   \title[Droplet propulsion onto gradient brushes without external energy supply]
  {Unidirectional Droplet Propulsion onto Gradient Brushes Without External Energy Supply}


\begin{document}

\begin{tocentry}

\includegraphics[width=8.25cm]{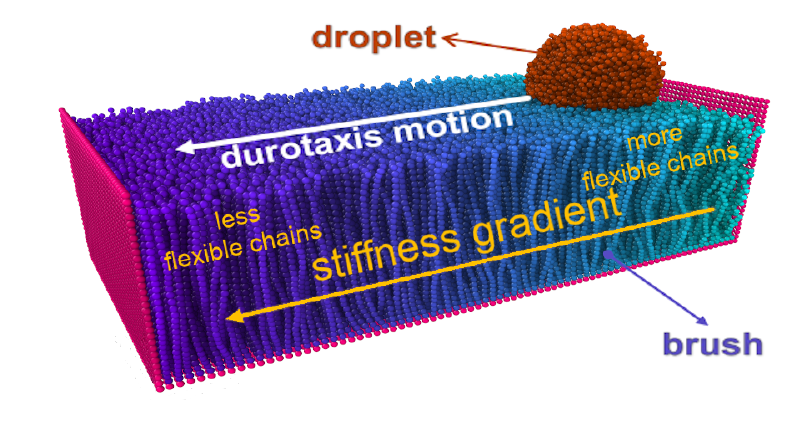}

\end{tocentry}

\begin{abstract}
Using extensive molecular dynamics simulation of 
a coarse-grained model, we demonstrate the 
possibility of sustained unidirectional motion (durotaxis) 
of droplets without external energy supply 
when placed on a polymer brush substrate
with stiffness gradient in a certain direction. 
The governing key parameters for the 
specific substrate design studied, which determine 
the durotaxis efficiency, are found to be the grafting 
density of the brush and the droplet adhesion to the brush surface,
whereas the strength of the stiffness gradient, the 
viscosity of the droplet or the length of the polymer chains
of the brush have only minor effect on the process. 
It is shown that this durotaxial motion 
is driven by the steady increase of the
interfacial energy between droplet and brush 
as the droplet moves from softer to stiffer parts
of the substrate whereby the mean driving force 
gradually declines with decreasing 
roughness  of the brush surface.  
We anticipate that our findings indicate
further possibilities in the area
of nanoscale motion without external energy supply.
\end{abstract}

\vspace{0.7in}

\section{INTRODUCTION}
The motion of nano-objects, for example, liquid nanodroplets,
can be provoked and sustained on solid substrates 
without an external energy supply. 
Moreover, the direction of motion can be controlled and
nanodroplets can move along predetermined trajectories. 
A way of achieving this effect is by placing the droplet onto
a \textit{gradient} substrate, that is, 
a substrate with a steadily varying property
along a specific direction. 
This is particularly attractive for the development of
various technologies in microfluidics, microfabrication, coatings,
nanoscale actuation and energy conversion, and biology \cite{Srinivasarao2001,Chaudhury1992,Wong2011,Lagubeau2011,Prakash2008,Darhuber2005, Yao2012, Li2018, Becton2016,vandenHeuvel2007,DuChez2019,Khang2015}.
Various possibilities for
the design of gradient substrates have been reported. 
For example, durotaxis
motion is caused by changes in stiffness along a substrate, as
has been shown in various natural processes in biology 
(\textit{e.g.}, cell movement on tissues) \cite{DuChez2019,Khang2015}
and in the case of real and \textit{in silico} experiments with liquid
droplets \cite{Theodorakis2017,Lo2000,Style2013,Chang2015,Pham2016,Lazopoulos2008,Becton2014,Barnard2015,Palaia2021,Tamim2021,Bardall2020}. 
Another characteristic example is the rugotaxis motion of 
droplets on wavy substrates with a gradient in the wavelength that 
characterises their pattern \cite{Theodorakis2022,Hiltl2016}.
Other possibilities include the use of wettability differences
\cite{Pismen2006,Wu2017} and
physical pinning \cite{Theodorakis2021}. 
Recent work has also highlighted the possibility of
uni-directional transport of small condensate droplets 
on asymmetric pillars\cite{Feng2020} or
three-dimensional capillary ratchets\cite{Feng2021}.
In the latter case, the motion can take place in one
or the other direction, depending on the surface tension
of the liquid. Other possibilities of directional motion
can take advantage of charge gradients that can achieve long-range
transport and are based on electrostatic\cite{Sun2019,Jin2022} or 
triboelectric charges.\cite{Xu2022}    
In contrast, motion caused by 
temperature gradient (thermotaxis) \cite{Zhang2022},
electrical current \cite{Dundas2009,Regan2004,Zhao2010,Kudernac2011},
charge \cite{Shklyaev2013,Fennimore2003,Bailey2008},
or even simple stretch \cite{Huang2014},
would require external energy
supply \cite{Becton2014}, as, also, in the case of
chemically driven droplets \cite{Santos1995,Lee2002},
droplets on vibrated substrates \cite{Daniel2002,Brunet2007,Brunet2009,Kwon2022}
or wettability ratchets \cite{Buguin2002,Thiele2010,Noblin2009,Ni2022}.

Inspired by our previous work with specific substrate designs 
that lead to the durotaxis\cite{Theodorakis2017} 
and rugotaxis\cite{Theodorakis2022} motion of nanodroplets
as motivated by the corresponding
experiments \cite{Style2013,Hiltl2016},
here, we propose a new design for the substrate, 
which is capable of sustaining the droplet motion. 
We consider a polymer brush, consisting
of polymer chains grafted onto a flat, solid surface, and 
the stiffness gradient is introduced to the brush
substrate by varying the stiffness of the polymer chains, 
which in practice 
amounts to tuning their persistence length.
To understand the mechanism of the durotaxis
motion on brush substrates and analyse the 
influence of relevant parameters for the brush
and the droplet (\textit{e.g.},
droplet adhesion to the substrate, droplet size,
viscosity, \textit{etc.}),
we have carried out extensive molecular dynamics (MD)
simulations of a coarse-grained (CG) model. 
This is crucial as the nanoscale motion of nano-objects
is usually controlled by tiny effects at the interface between
the droplet and the substrate resulting from the 
molecular interactions between the two, which only
a method with molecular scale resolution can capture.
As in the case of durotaxis\cite{Theodorakis2017}
and rugotaxis\cite{Theodorakis2022} motions, we
find that the motion is caused by a gradient in
the droplet--substrate interfacial energy, which
translates into an effective force that drives
the droplet towards the stiffer, flatter parts of the 
substrate. Moreover, we find that the 
efficiency of the durotaxis motion for
brush substrates is higher for moderate values
of the grafting density and droplet adhesion to the
substrate as well as for smaller droplets and longer brush
chains. Surprisingly, we have not observed a significant
effect of the stiffness gradient in the case of
brush substrates when the motion was successful.
We anticipate that our study will shed some light 
into the durotaxis motion of droplets on brush, gradient
substrates, thus providing further possibilities in
nanoscale science and technology \cite{Barnard2015},
for various medicine and engineering 
applications \cite{Barthlott2016,Khang2015}. Moreover,
brush substrates share connection with various biological
surfaces that expel various exogenous substances from
their structure,\cite{Badr2022} such as the the mucus layer from airway 
epithelia,\cite{Button2012} while the gradient concept
plays an important role in applications of 
regenerative medicine.\cite{Khang2015}
In the following, we discuss our simulation model and
methodology. Then, we will present and discuss our results,
and in the final section we will draw our conclusions.

\section{MATERIALS AND METHODS}

\begin{figure}[bt!]
\centering
 \includegraphics[width=0.75\columnwidth]{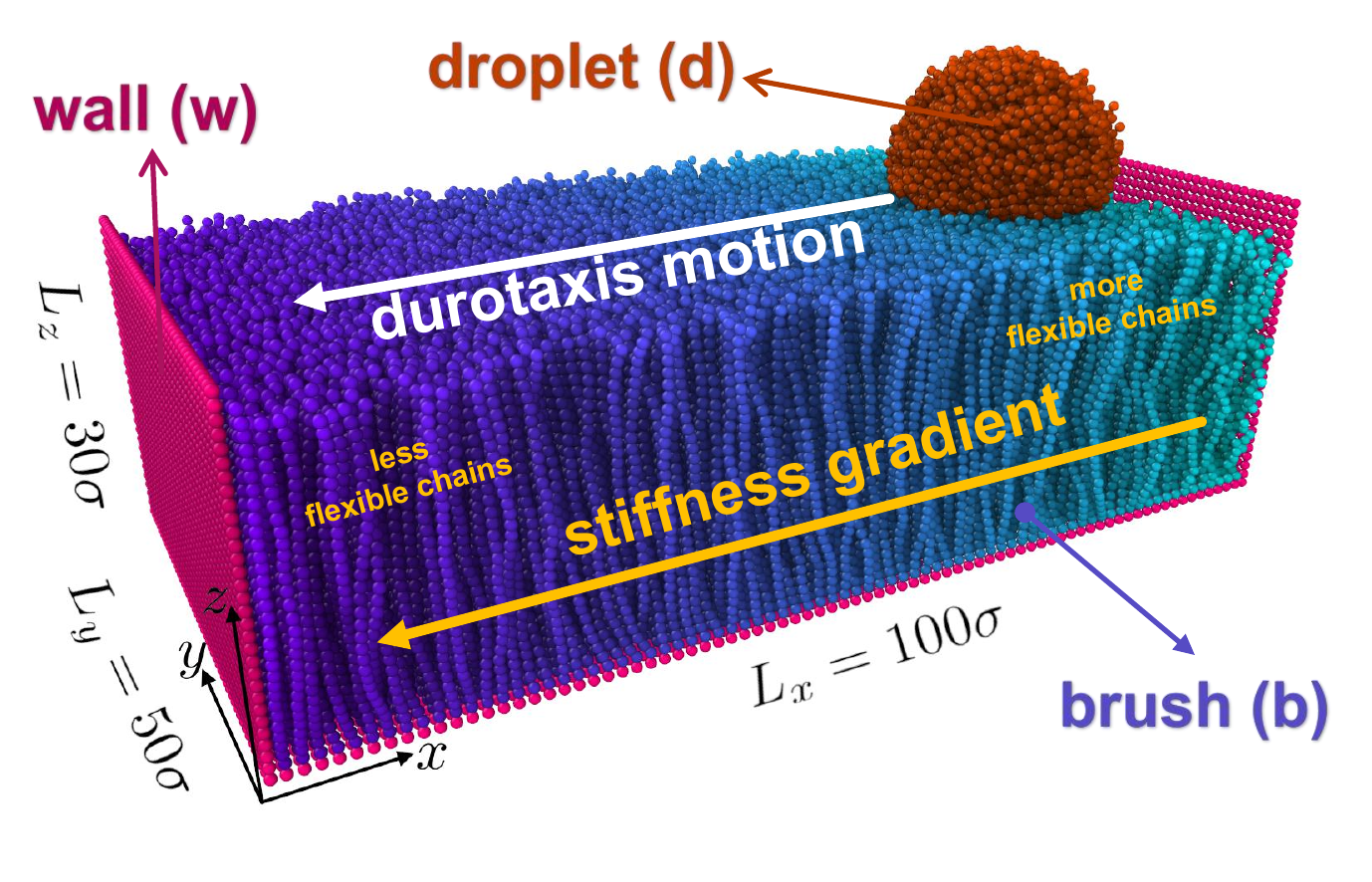}
\caption{\label{fig:1} Typical initial configuration of 
the system, where the droplet is placed on the softest
end of the brush substrate.
Here, $N_{\rm b}=30$, $N=4000$, and $N_{\rm d}=10$ beads,
$\sigma_{\rm g}=0.6~\sigma^{-2}$, and 
$\varepsilon_{\rm db}=0.6~\epsilon$.
At the softest end $k_{\rm \theta}=0~\epsilon/{\rm rad}^2$,
while at the stiffest end
$k_{\rm \theta}=80~\epsilon/{\rm rad}^2$, 
with a linear gradient of $k_{\theta}$ between them
in the $x$ direction.
$L_{x}$, $L_{y}$, and $L_{z}$, indicate
the dimensions of the immobile walls.
See text for further details.
The snapshot of the system was obtained using 
Ovito software.\cite{Stukowski2010} }
\end{figure}

Our system consists of a polymer-brush substrate
and a droplet placed on its soft part (Fig.~\ref{fig:1}). 
We have found that the durotaxis motion in the case
of brush substrates takes place from the
softer towards its stiffer parts,
which is in line with previous simulation findings
for another substrate design with stiffness 
gradient.\cite{Theodorakis2017}
In the direction of the
stiffness gradient, the substrate has length, 
$L_x=100~\sigma$ ($\sigma$ is the unit of length), 
while in the $y$-direction, $L_y=50~\sigma$,
which guarantees that mirror images of the droplet in this
direction will not interact during the course of the
simulation due to the presence of periodic boundary 
conditions that are applied in all Cartesian directions.
Finally, two walls are placed normal to the
$x$ direction as shown in Fig.~\ref{fig:1} and
the size of the box in the $x$ direction is large
enough to guarantee that there are no interactions
between the walls or the polymers on the two opposite sides
of the simulation domain in the $x$ direction.
Wall beads were kept immobile during the
simulation. 

The standard bead--spring model \cite{Kremer1990} has
been employed in our simulations. In this model,
interactions between any of the system components, 
\textit{i.e.} the drop (d), the brush (b), and the wall (w) beads,
are described by means of the Lennard-Jones (LJ) potential
\begin{equation}\label{eq:LJpotential}
U_{\rm LJ}(r) = 4\varepsilon_{\rm ij} \left[  \left(\frac{\sigma_{\rm  ij}}{r}
\right)^{12} - \left(\frac{\sigma_{\rm ij}}{r}  \right)^{6}    \right],
\end{equation}
where $r$ is the distance between any pair of beads in the system
within a cutoff distance.
Indices ${\rm i}$ and ${\rm j}$ in Eq.~\ref{eq:LJpotential}
indicate the type of beads. The size of the beads
is $\sigma_{\rm ij} = \sigma$ for all interactions. The LJ potential
is cut and shifted at the cutoff distance, $r_{\rm c}=2.5~\sigma$,
for the interaction between the droplet (d) beads, 
as well as the interaction between the droplet and
the brush (b) beads. In contrast, a purely repulsive potential for 
the interaction between the brush beads, as well
as between the brush and the wall (w) beads, was considered,
that is, in this case, $r_{\rm c}=2^{1/6}~\sigma$.
The strength of the attractive interactions is determined
by the parameter $\varepsilon_{\rm ij}$
of the LJ potential \cite{Theodorakis2011} .
In our study, $\varepsilon_{\rm dd}=1.5~\epsilon$, with
$\epsilon$ defining the energy scale. 
Moreover, $\varepsilon_{\rm bb}=\epsilon$, while
$\varepsilon_{\rm db}$ is the parameter
that controls the attraction (adhesion) of the droplet
to the substrate beads and in our
study ranged from $0.1$ to $1.2~\epsilon$. 

The grafting density, $\sigma_{\rm g}$, 
is varied from $0.1$ to $1.0~\sigma^{-2}$ in our study. 
The size of the droplets can also vary
through the total number of beads that the droplet 
contains, which ranged between $2\times10^3$ and $1.6\times10^4$ 
beads in our simulations.
These beads belong to fully flexible, linear polymer chains. 
By varying the length of the droplet chains, e.g., from 
$10$ to $80$ beads, we can alter the viscosity of the 
droplet.\cite{Theodorakis2017}
The finite extensible nonlinear elastic (FENE) 
potential \cite{Kremer1990} was used to tether together 
consecutive beads in these polymer chains,
as well as the polymer beads along the linear brush-polymer
chains.
The mathematical expression for the FENE potential is as follows:
\begin{equation}\label{eq:KG}
 U_{\rm FENE}(r) = -0.5 K_{\rm FENE} R_{\rm 0}^2 \ln \left[ 1 - \left(\frac{r}{R_{\rm 0}} \right)^2  \right],
\end{equation}
where $r$ is the distance between two 
consecutive beads along the polymer backbone, while
$R_{\rm 0}=1.5~\sigma$ expresses 
the maximum extension of the bond, and
$K_{\rm FENE} = 30~\epsilon/\sigma^2$
is an elastic constant. 
Lengths of the polymer chains in the droplet 
greater than $N_{\rm d}=10$ 
guarantee that there are no evaporation effects
and the vapour pressure is hence sufficiently 
low \cite{Tretyakov2014} .
We have also investigated the effect of the length, $N_{\rm b}$,
of the polymer chains of the brush on the durotaxis motion, 
by choosing different lengths, 
namely $N_{\rm b}=15$, $30$, and $50$ beads.

The stiffness gradient is imposed on the brush substrate
by varying the stiffness of the individual brush polymer-chains. 
The total length of the brush chains, $N_{\rm b}$, was 
the same for all chains,
but their stiffness
changed depending on the Cartesian coordinate
of their grafting site in the $x$ direction, 
\textit{i.e.} chains with the same position $X$
of their grafted end have the same stiffness.
The chain stiffness was controlled by using a harmonic angle
potential for every triad of consecutive beads along the 
polymer chain and tuning its strength through the harmonic
constant $k_{\theta}$. 
The form of the harmonic potential reads:
\begin{equation}\label{eq:harmonice_angle}
 U_{\theta_{\rm ijk}}(\theta) = k_{\theta} (\theta_{\rm ijk} - \theta_0)^2,
\end{equation}
where $\theta_{\rm ijk}$ is the angle between three
consecutive beads $\rm i$, $\rm j$, and $\rm k$ along
a brush polymer chain and, $\theta_0=\pi~{\rm rad}$,
is the equilibrium angle. 
A linear gradient in the stiffness constant, $k_{\theta}$, is
considered in our study to explore the properties
of our systems. As we will discuss later, while
the gradient in the stiffness of the substrate is 
necessary to initiate and maintain the durotaxis motion,
the system is rather insensitive to the exact value
of the gradient and the key parameters for the brush 
substrate turn out to be the grafting density, 
$\sigma_{\rm g}$, and the substrate wettability as
controlled via the parameter $\varepsilon_{\rm db}$.
The reasons for this will be revealed 
during the discussion of our results.

To evolve our system in time, the 
Langevin thermostat was used, whose details have been discussed 
in previous studies\cite{Theodorakis2010a,Theodorakis2010b} .
Hence, the simulations are in practice realised 
in the canonical ensemble,\cite{Schneider1978}
where the temperature, $T$, of the system 
fluctuates around a predefined value
$T=\epsilon/k_{B}$,  with $k_{B}$ being the 
Boltzmann constant and $\epsilon$ the energy unit.
For the integration of the equations of motion, 
the LAMMPS package \cite{Plimpton1995} was employed. 
The MD time unit is $\tau =\sqrt{m\sigma^2/\epsilon}$,
where $m$ is the unit of mass,
and the integration time step was $\Delta t =0.005~\tau$. 
Typical simulation trajectories start from configurations 
like the one presented in Fig.~\ref{fig:1} with the total
length of each trajectory being $10^8$ MD integration steps.
If a droplet fully transverses the substrate 
from the softest to the stiffest end of the substrate,
then the durotaxis motion is considered as successful. 
To ensure reliable statistics an ensemble of ten
independent trajectories
with different initial conditions (by changing the
initial velocities assigned to each particle) was used 
for each set of system parameters. Our results are based on
the analysis of these trajectories for each case.

\section{RESULTS AND DISCUSSION}

\begin{figure}[bt!]
\centering
 \includegraphics[width=\columnwidth]{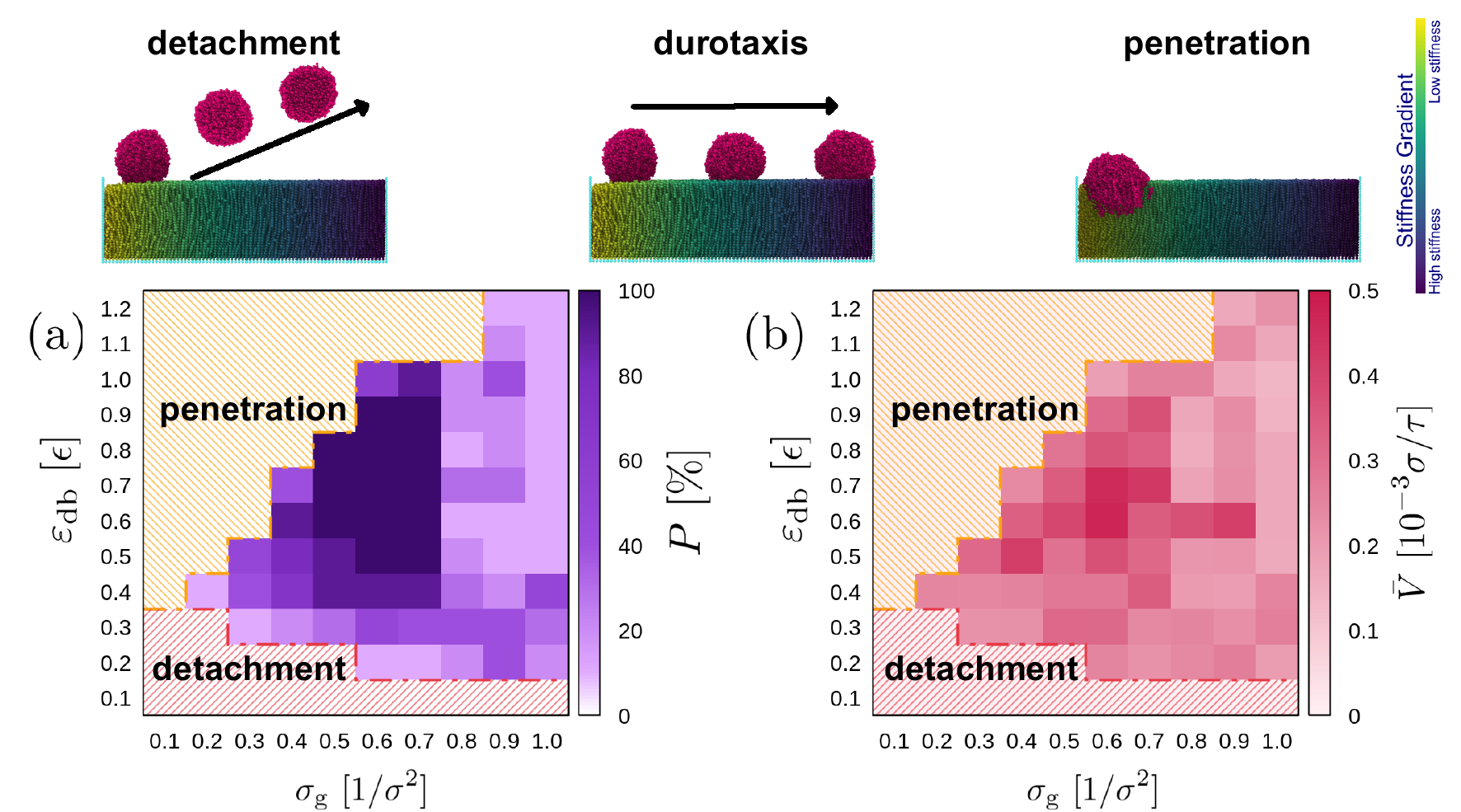}
\caption{\label{fig:2} (a) Regime map indicating the
probability, $P$ (color scale),
that a droplet will cover the full distance
over the substrate in the $x$ direction from the softest
to the stiffest part (successful durotaxis cases) 
for different values of the droplet--substrate
attraction, $\varepsilon_{\rm db}$, and the 
grafting density, $\sigma_{\rm g}$. 
Probabilities, $P$, are based on an ensemble of 
ten independent simulations for each set of parameters. 
The regimes where the
droplet penetrates into the brush or detaches from the
substrate due to the weak $\varepsilon_{\rm db}$ attraction
are also shown with a different color. 
(b) The color map indicates the average
velocity of the droplet, $\bar{\varv}=L'_x/t$, for the successful 
durotaxis cases, where $t$ is the time that the droplet
needs to cross the full length of the brush substrate
in the $x$ direction, and $L'_x$ is the actual
distance covered by the centre-of-mass of the droplet
for each successful case. 
$N=4000$, $N_{\rm d}=10$, $N_{\rm b}=30$ beads. The 
stiffness constant for the polymer chains in the
softest part of the substrate 
is zero (fully flexible chains), growing linearly to
$k_{\rm \theta}=80~\epsilon/{\rm rad}^2$ at the stiffest 
part of the substrate. Since $L_{x}=100~\sigma$,
the stiffness gradient is 
$\Gamma=0.8~\epsilon/{\rm rad}^2\sigma$.
Snapshots on top of the plot indicate examples of 
detachment, durotaxis, and penetration, as indicated.
}
\end{figure}

\begin{figure}[bt!]
    \centering
    \includegraphics[width=0.5\textwidth]{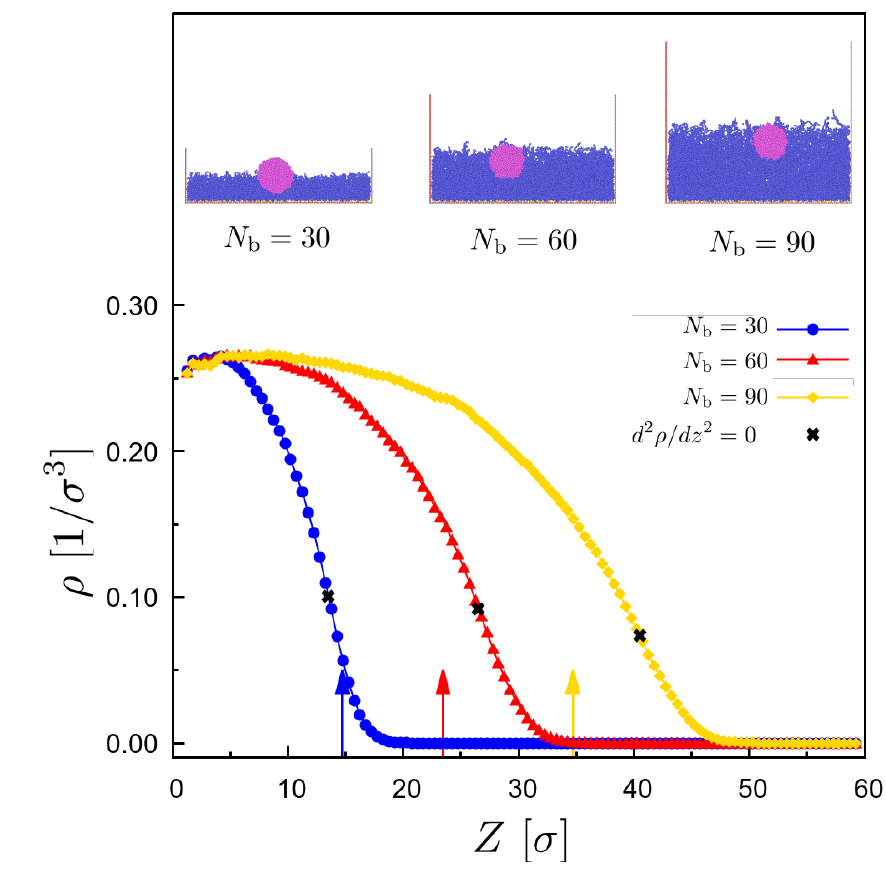}
    \caption{ \label{fig:3} Density profile in the $z$ direction
    for three polymer brushes at $\sigma_g = 0.6~\sigma$
    with fully flexible chains of length, $N_{\rm b}=30$, $60$, and $90$ 
    beads, as indicated. The inflection point of the curves, 
    $d^2\rho/dz^2=0$, marked with {\color{black}$\boldsymbol{\times}$},
    is shown for each case, which corresponds to the height of
    the brush. The position of the centre of mass of the droplet
    in the $z$ direction is marked with arrows of the same 
    colour for each case. Here, $N=4000$ and $N_{\rm d}=10$ beads, and $\varepsilon_{\rm db}=0.6~\epsilon$.
    Snapshots for each case are shown
    in the plot. In the case of brushes with longer chains, the
    droplet is immersed deeper into the brush.
 }
\end{figure}

By exploring a wide range of parameters, we have found that 
the grafting density, $\sigma_{\rm g}$, and
the attraction strength between the
droplet and the substrate, $\varepsilon_{\rm db}$,
are two key parameters of the substrate design, 
since they greatly affect the possibility
for successful durotaxis.
Figure~\ref{fig:2} presents the regime maps as
a function of these two parameters with the
probability, $P$, of successful durotaxis 
calculated from the ensemble of ten independent
simulations for each set of parameters.
However, in our \textit{in silico} experiments,
apart from durotaxis motion, we have also
documented situations in which the droplet penetrates
into the substrate or detaches from it. 
Our results indicate that small
attraction strengths will lead to droplet
detachment. 
This is more probable at smaller grafting densities due to fewer 
brush--droplet interactions. 
In contrast large values
of the attraction strength, $\varepsilon_{\rm db}$,
can lead to the penetration of the brush substrate
by the droplet, again for smaller values of the grafting 
density, $\sigma_{\rm g}$.
Moreover, our results suggest that the difficulty of 
the droplet to penetrate into the substrate
increases roughly linearly with the 
grafting density up to $\varepsilon_{\rm db} = \epsilon$, 
as evidenced by the linear boundary in the regime maps.
In fact, penetration becomes impossible when
$\sigma_{\rm g} \geq 0.9~\sigma^{-2}$, since there is not
enough space among the brush beads to accommodate 
additional droplet beads. 
Moreover, the degree of brush penetration also
depends on the length, $N_{\rm b}$, of the 
brush chains, as shown in Fig.~\ref{fig:3}.
In general, our data suggest that droplets
are immersed deeper in brushes
with longer chains, as shown here in the
case of fully flexible polymers and judging
by the centre-of-mass of the droplet with
respect to the position of the brush surface,
as defined by the inflection point at the
density profile (Fig.~\ref{fig:3}).
A more detailed study on this effect could potentially
reveal more details for droplets
immersed in brush substrates, but this clearly
goes beyond the scope of our current study.

When penetration and detachment of the droplet are avoided, 
then persistent durotaxis motion is observed
with a certain probability, $P$, which depends on
the choice of $\sigma_{\rm g}$, 
and $\varepsilon_{\rm db}$ (Fig.~\ref{fig:2}a).
From the results of Fig.~\ref{fig:2}a, we find that the range
$0.5~\sigma^{-2} \leq \sigma_{\rm g} \leq 0.7~\sigma^{-2}$
combined with 
$0.5~\epsilon \leq \varepsilon_{\rm db} \leq 0.9~\epsilon$,
in general, provides certainty in the success 
of the durotaxis motion ($P=100~\%$),
since all our droplets were able to fully cross the substrate
from the softest to the stiffest parts of the brush within
the available simulation time of $10^8$ MD time steps. 
As the grafting density increases above $0.7~\sigma^{-2}$,
however, we observe that the
probability of durotaxis success suddenly decreases.
In this case, a higher density of the brush chains
increases the resultant brush stiffness
owing to the close packing of the chains, which
leads to a situation that the role of the 
nominal stiffness of the individual brush chains, $k_{\theta}$, in 
determining the effective stiffness gradient becomes negligible.
As we will see later in our discussion concerning the 
underlying durotaxis mechanisms, the extent of disarray 
of the brush chain end-monomers 
at the brush surface is also reduced pointing to 
a rather flat density profile.
 
To determine the efficiency of the durotaxis motion, 
we have computed the average velocity, $\bar{\varv}$, of 
the droplet for the successful cases for each set of parameters
$\sigma_{\rm g}$ and $\varepsilon_{\rm db}$ (Fig.~\ref{fig:2}b). 
Our results indicate that
the probability, $P$, for success rather correlates with
the highest average velocity, $\bar{\varv}$, but
large values of $\bar{\varv}$ can also be obtained
in certain cases where $P<100~\%$, for example, the case
$\sigma_{\rm g} = 0.9~\sigma^{-2}$,
$\varepsilon_{\rm db} \sim 0.6~\epsilon$ (Fig.~\ref{fig:2}b).
This is a clear indication that durotaxis motion is
controlled by tiny effects that can greatly 
influence the outcome of the experiments. 
Moreover, obtaining reliable statistics in cases of
$P<100\%$ remains a challenge in MD since this would
require the realisation of a large number of simulations.
Hence, as $P \longrightarrow 0$ obtaining reliable statistics
becomes more of a challenge and outliers in the statistics
are more probable.
In summary, the plots of Fig.~\ref{fig:2} suggest that
if one would like all droplets
to fully cross the substrate in the direction of 
the stiffness gradient in the shortest time, 
then values of $\sigma_{\rm g} = 0.6~\sigma^{-2}$
and $\varepsilon_{\rm db} \approx 0.7~\epsilon$
would constitute an optimal choice in the \textit{in silico}
experiments.
Hence, we argue that moderate values of $\sigma_{\rm g}$ 
and $\varepsilon_{\rm db}$ favour successful and
efficient (in terms of time to cross the whole substrate)
durotaxis motion.

\begin{figure}[bt!]
\centering
 \includegraphics[width=\columnwidth]{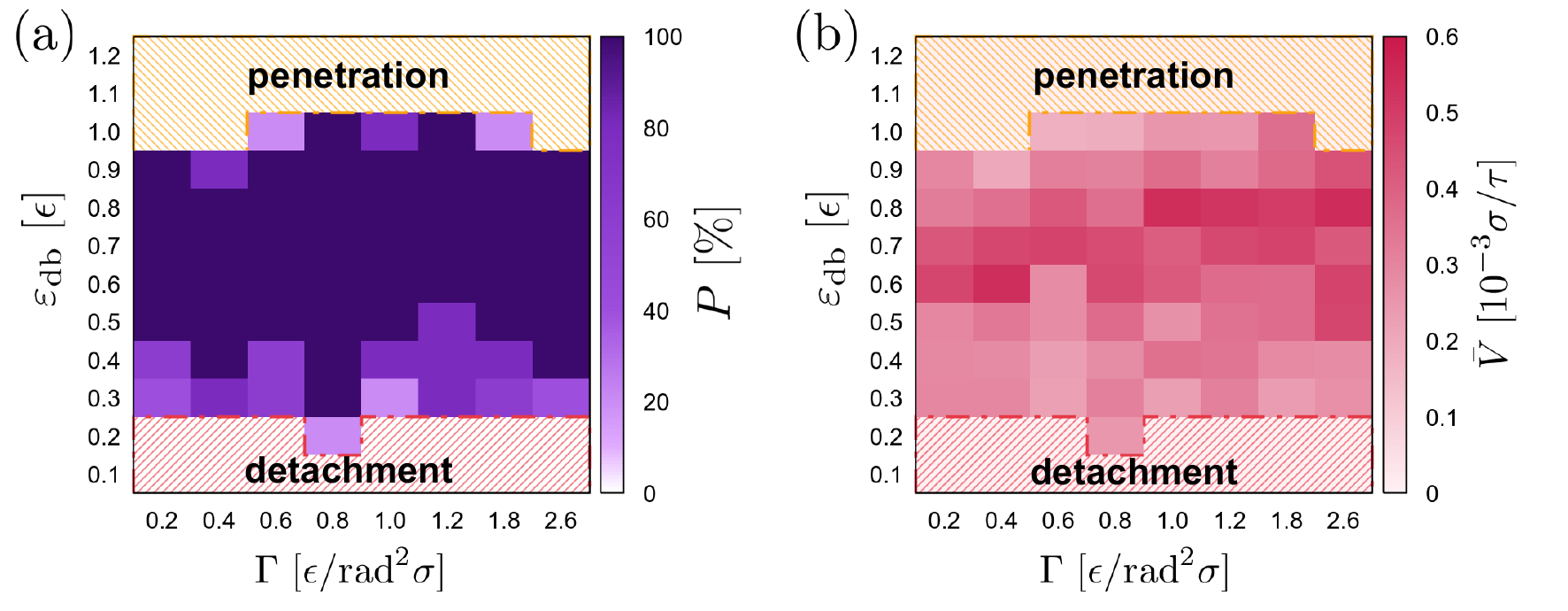}
    \caption{\label{fig:4} (a) Regime map indicating the
    probability, $P$ (color scale), that a droplet will
    cover the full distance over the substrate in the $x$ 
    direction from the softest to the stiffest part
    (sucessful durotaxis cases) for different values 
    of the droplet--substrate attraction, $\varepsilon_{\rm db}$, 
    and the stiffness gradient $\Gamma=dk_\theta/dx$. 
    The regimes where the droplet penetrates into the brush 
    or detaches from the substrate due to
    the weak $\varepsilon_{\rm db}$ attraction 
    are also shown with a different color. 
    (b) The color map indicates the average
velocity of the droplet, $\bar{\varv}=L'_x/t$, for the successful 
durotaxis cases, where $t$ is the time that the droplet
needs to cross the full length of the brush substrate
in the $x$ direction, and $L'_x$ is the actual
distance covered by the centre-of-mass of the droplet
for each successful case.  The 
stiffness constant, $k_{\theta}$, for the polymer chains in the
softest part of the substrate 
is zero (fully flexible chains), growing linearly to
its highest value at $x=L_{x}=100~\sigma$,
which depends on the chosen stiffness gradient, $\Gamma$.
Here, $\sigma_{\rm g} = 0.6~{\sigma^{-2}}$, 
$N=4000$, $N_{\rm b}=30$, and $N_{\rm d}=10$ beads. 
}
\end{figure}

Finally, we have explored the effect of the
stiffness gradient on the durotaxis motion. 
Here, we have picked the best case of Fig.~\ref{fig:2},
that is $\sigma_{\rm g} = 0.6~\sigma^{-2}$ and 
$\varepsilon_{\rm db} = 0.6~\epsilon$, and
varied the stiffness gradient $\Gamma=dk_\theta/dx$
in the range 0.2--2.6 for fully flexible chains
at the softest part of the substrate, which
has actually provided the best result in
terms of durotaxis success and efficiency.
Overall, we have found that
durotaxis is insensitive to the value of $\Gamma$
in the range $0.5~\epsilon \leq \varepsilon_{\rm db} \leq 0.9~\epsilon$
(Fig.~\ref{fig:4}a), in contrast to what
has been observed for other \textit{in silico} 
substrate designs\cite{Theodorakis2017}. 
Moreover, the average velocities are spread
out with small variations and no indication
of a clear trend (Fig.~\ref{fig:4}b) that would indicate
that a larger stiffness gradient would lead to more efficient
durotaxis motion exists, which has been the case for other
substrate designs\cite{Theodorakis2017}.
Moreover, since the motion
is most efficient when the softest part
consists of fully flexible chains ($k_{\theta}=0~\epsilon/{\rm rad}^2$)
might suggest that brush substrates with polymer chains 
of small persistence lengths (as soft as possible) are more suitable
for successful durotaxis motion. Henceforth, all our 
results refer to brush substrates with fully
flexible chains at their softest part and 
stiffness gradient $\Gamma=0.8~\epsilon/{\rm rad}^2\sigma$.

\begin{figure}[bt!]
\centering
 \includegraphics[width=\columnwidth]{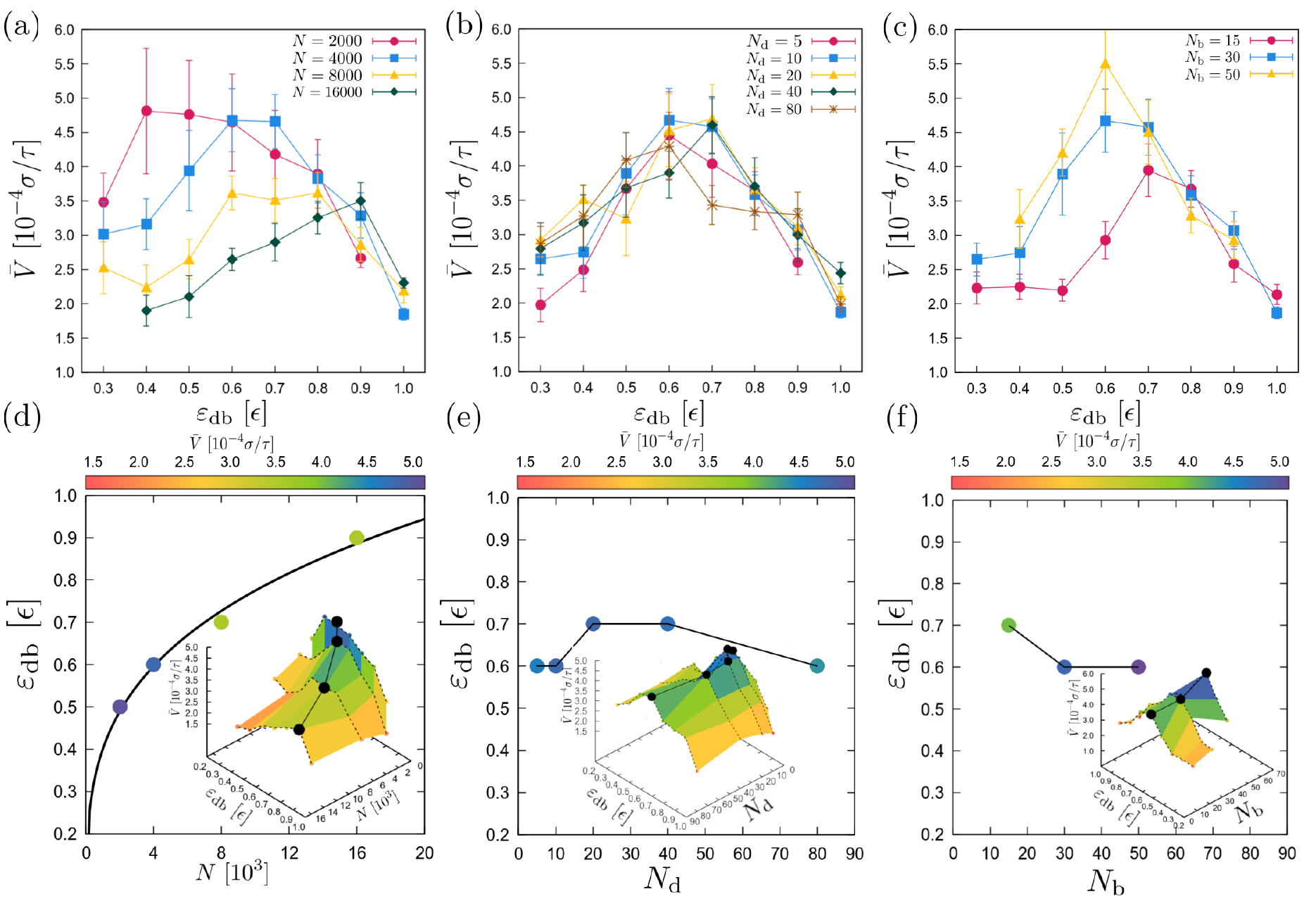}
\caption{\label{fig:5} Average
droplet velocity, $\bar{\varv}$, as a function of
$\varepsilon_{\rm db}$ for different (a) droplet size,
$N$ ($N_{\rm d}=10$, $N_{\rm b}=30$ beads),
(b) chain length of the polymer chains of the droplet, 
$N_{\rm d}$ ($N=4000$ and $N_{\rm b}=30$ beads) , and 
(c) brush polymer chain length, $N_{\rm b}$
($N=4000$ and $N_{\rm d}=10$ beads), as indicated.
(d) Documented maximum average velocity, $\bar{\varv}$,
indicated by the colour map, as a function of
$\varepsilon_{\rm db}$ and $N$. 
Inset shows the average velocity, $\bar{\varv}$, 
for all pairs of ($\varepsilon_{\rm db}$, $N$). 
The black points indicate the pairs
($\varepsilon_{\rm db}$, $N$) for which we have
the maximum average velocity, $\bar{\varv}$,
which is shown in the main plot.
(e) Same as (d), but data are plotted as
a function of ($\varepsilon_{\rm db}$, $N_{\rm d}$).
(f) In this case data are plotted as a function of
($\varepsilon_{\rm db}$, $N_{\rm b}$).
$\sigma_{\rm g} = 0.6~\sigma^{-2}$,
$\Gamma=0.8~\epsilon/{\rm rad}^2\sigma$ in all cases.
Panels (a-c) here show data as a function of $\varepsilon_{\rm db}$ with one other quantity varying, the others held constant. Therefore, it does
take care to carry out a systematic study with as much held constant between data as useful. Panels (d-f) visualize the
same data as in the top panels but with a different visualization in terms of maximum velocities.
}
\end{figure}

In the following, we examine the effect of various parameters
on the efficiency of the durotaxis motion. For our analysis,
we have picked the case $\sigma_{\rm g} = 0.6~\sigma^{-2}$,
which has shown the best performance in terms of the
probability, $P$, and the average velocity, 
$\bar{\varv}$, in our study
(Fig.~\ref{fig:2}) and therefore would most probably allow for
exploring a wider range of the parameter space.
Then, Figs~\ref{fig:5}a,d illustrate the
dependence of the average velocity, $\bar{\varv}$,
on the attraction strength, $\varepsilon_{\rm db}$,
for various droplet sizes, $N$. 
We observe that the fastest durotaxis motion
takes place for the smallest droplets, namely
$N=2000$ beads. As $\varepsilon_{\rm db}$ increases, 
a maximum value of the average velocity, $\bar{\varv}$, 
appears for each case with droplets of different size. 
In addition, we consistently see that this maximum value
of $\bar{\varv}$ becomes smaller for larger droplets.
For example, for the droplet of
$N=2000$ beads the average velocity measured over
ten trajectories is 
$\bar{\varv} \approx 4.7\times10^{-4}~\sigma/\tau$,
while $\bar{\varv} \approx 3.5\times10^{-4}~\sigma/\tau$
for droplets with $16000$ beads. These differences
are generally considered small, especially compared
to other \textit{in silico} experiments.\cite{Theodorakis2017,Theodorakis2022}
Moreover, we see that the maximum shifts to higher
values of the attraction strength, $\varepsilon_{\rm db}$,
as the size of the droplet increases.
For example, the maximum velocity is observed when 
$\varepsilon_{\rm db}=0.4~\epsilon$ for droplets
with $N=2000$ beads and for 
$\varepsilon_{\rm db}=0.9~\epsilon$ for droplets of
$16000$ beads (Fig.~\ref{fig:5}d).
Finally, for small droplets we observe a steep
increase of the average velocity
with $\varepsilon_{\rm db}$,
and then a slow decrease (Fig.~\ref{fig:3}a). 
For medium-size droplets,
(\textit{i.e.} $N=4000$ and $N=8000$ beads),
there is a smooth maximum that develops in the middle
range of $\varepsilon_{\rm db}$, \textit{i.e.}
$0.6~\epsilon \leq \varepsilon_{\rm db} \leq 0.7~\epsilon$, 
while in the case of droplets with $N=16000$ beads
there is a maximum that slowly develops as 
$\varepsilon_{\rm db}$ increases, which is followed
by a steeper decrease when $\varepsilon_{\rm db} > 0.9~\epsilon$.
In summary, we observe that the size of the droplets
is an important parameter for the durotaxis motion.

The next parameter to examine is the chain length, 
of the polymers comprising the droplet,
$N_{\rm d}$ (Figs~\ref{fig:5}b, e).
In practice, longer chain lengths would result
in a larger droplet viscosity. 
During this and previous work with 
polymer liquid-droplets\cite{Theodorakis2017}
we have determined that the most relevant values for
our study are within the range $10 \leq N_{\rm d} \leq 80$ beads.
Interestingly, the droplet viscosity seems not to
play an important role in the overall efficiency
of the durotaxis motion, for a given
attraction strength, $\varepsilon_{\rm db}$.
In other words, droplets with different $N_{\rm d}$
would exhibit a similar durotaxial efficiency for a 
specific choice of $\varepsilon_{\rm db}$. 
As a result, the maximum average velocity, 
$\bar{\varv}$, as a function of $\varepsilon_{\rm db}$
appears at
$\varepsilon_{\rm db} \sim 0.6-0.7~\epsilon$ (Fig.~\ref{fig:5}e).
Hence, we can conclude that droplets with
different viscosity will have a similar
durotaxis performance and, here, a moderate choice for
the value of droplet--substrate attraction strength
would yield the fastest durotaxis motion.

The effect of the length of the brush chains,
$N_{\rm b}$, on the durotaxis motion is 
shown in Figs~\ref{fig:5}c, f.
We find that the
larger the length, $N_{\rm b}$, the more efficient 
the durotaxis motion becomes. Although a
larger difference is noticed when $N_{\rm b}$
was doubled from 15 to 30 beads, a saturation
in our data occurred when $N_{\rm b}$
increased from 30 to 50 beads. 
Overall, our results indicate that brushes with longer polymer
chains favor the durotaxis motion. 
Moreover, for $N_{\rm b}=30$ and 50 beads, the maximum
velocity is found for $\varepsilon_{\rm db} = 0.6~\epsilon$,
and hence is independent of the choice of $N_{\rm b}$.
The reasons for this behavior will become more 
apparent as we will
discuss the underlying mechanism of the droplet motion 
in the following.

\begin{figure}[bt!]
\centering
 \includegraphics[width=1.0\columnwidth]{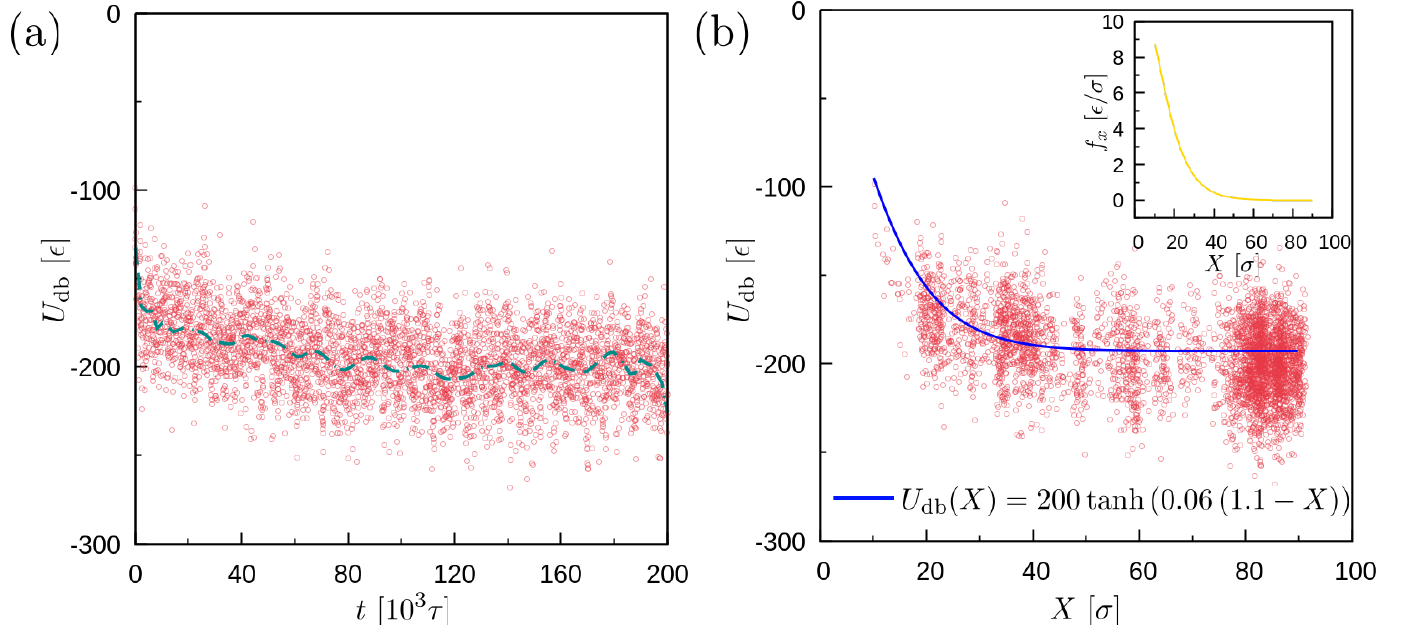}
\caption{\label{fig:6} Droplet--brush interfacial energy,
$U_{\rm db}$, as a function of time, $t$ 
(a, dashed line is a guide for the eye), and the
position, $X$, of the centre of mass of the droplet in
the $x$ direction (b), for a case with successful durotaxis
($N=4000$, $N_{\rm d}=10$, and $N_{\rm d}=30$ beads. 
$\varepsilon_{\rm db}=0.6~\epsilon$ and $\sigma_{\rm g}=0.6~\sigma^{-2}$).
Inset shows the force, $f_x=-\frac{\partial U_{\rm db}}{\partial x}$ 
based on a nonlinear fit of the $\tanh$ function
on the $U_{\rm db}$ data. The higher concentration of points  in
panel (b) at certain ranges of $X$ simply indicates that 
the droplet spends more time 
at these positions as it moves to the stiffest parts of the substrate.
The fit function of the $U_{\rm db}$ only provides 
an average picture of the decay
of the interfacial energy.
}
\end{figure}

\begin{figure}[bt!]
    \centering
    \includegraphics[width=1.0\columnwidth]{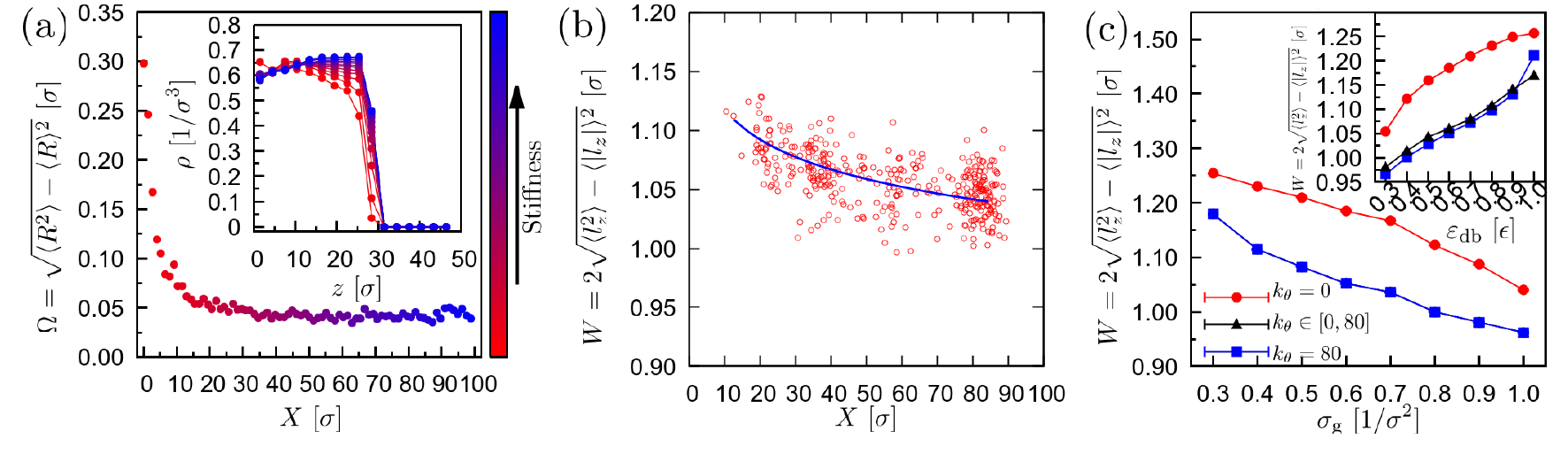}
    \caption{\label{fig:7} 
    (a) The standard deviation in the end-to-end distance
    of the brush polymer chains 
    as a function of their grafting position,
    $X$. Larger values of $X$ correspond to the stiffer parts of
    the substrate. Inset shows the density profile in the
    $z$ direction at different positions, $X$. The colour reflects
    the stiffness of the chains ($k_{\theta}$).
    $\varepsilon_{\rm db}=0.6~\epsilon$, 
    $\sigma_{\rm g}=0.6~\sigma^{-2}$.
    $\Gamma=0.8 ~\epsilon/{\rm rad}^2\sigma$ and 
    initial stiffness $0~\epsilon/{\rm rad}^2$ at the 
    softest end. $N=4000$, $N_{\rm d}=10$,
    $N_{\rm b}=30$ beads.
    (b) Interpenetration length, $W$. 
    $l_z$ is the average distance of
    contact pairs between the droplet and the brush
    for the same set of parameters as in (a) 
    with $X$ here indicating the centre-of-mass position
    of the droplet in the $x$ direction along the gradient. 
    (c) Average interpenetration length as a function of the
    grafting density, $\sigma_{\rm g}$, for substrates with
    constant stiffness ($k_{\theta}=0~\epsilon/{\rm rad}^2$ or 
    $k_{\theta}=80~\epsilon/{\rm rad}^2$) or with stiffness 
    gradient $\Gamma=0.8 ~\epsilon/{\rm rad}^2\sigma$ and 
    initial stiffness $0~\epsilon/{\rm rad}^2$ at the 
    softest end, as indicated. $N=4000$, $N_{\rm d}=10$,
    $N_{\rm b}=30$ beads. $\varepsilon_{\rm db}=0.6~\epsilon$ in
    the main plot and $\sigma_{\rm g}=0.6~\sigma^{-2}$ in the
    inset.
    }
    
\end{figure}

From our earlier studies, we have seen that 
the minimization of the interfacial energy, $U_{\rm db}$,
between the droplet and the substrate is the driving force
for the durotaxis motion in the case of substrates
with stiffness gradient \cite{Theodorakis2017} or
wrinkle substrates with a gradient in the wavelength
characterising the wrinkles \cite{Theodorakis2022}. This
has also been found in the case of a nanoflake 
on substrates with stiffness gradient.\cite{Chang2015}
Moreover, we have argued that the efficiency of 
the motion depends on 
the rate of change of the interfacial energy along the
gradient direction.
Hence, it is relevant to examine the interfacial
energy, $U_{\rm db}$, as a function of time, $t$, and
the coordinate, $X$, of the centre-of-mass of the droplet
along the substrate (see Fig.~\ref{fig:6} 
showing these quantities for a successful
durotaxis case). 
Our results indicate that $U_{\rm db}$ over time reaches
more negative values (Fig.~\ref{fig:6}a), 
which corresponds to a larger number of attractive pair
interactions between the droplet and the 
substrate. Importantly, we also see that the
energy decreases as a function of the 
position of the droplet, $X$, which
clearly indicates that the droplet
moves to areas of more negative energy (Fig.~\ref{fig:6}b)
towards the stiffer parts of the substrate (increasing $X$).
A larger decrease of the energy takes place
in the initial soft parts of the substrate
at small $X$ values, while the change in
the substrate--droplet interfacial energy is smaller
after the droplet has moved a distance about
40~$\sigma$. Moreover, we can see that the
droplet does not spend the same time at each
position $X$, as seen by the density of points 
at certain positions, $X$, and knowing that
samples have been taken over time at equal
intervals. Hence, the durotaxis motion of
the droplet cannot be characterised as a steady
state. To illustrate the effect of the gradient
in the interfacial energy, $U_{\rm db}$, one can actually plot
the negative derivative of the interfacial energy
(inset of Fig.~\ref{fig:6}b) after
performing a suitable fit on the $U_{\rm db}$. 
This derivative
would correspond to an on-average driving force
$f_x=-\frac{\partial U_{\rm db}}{\partial x}$ that propels
the droplet toward the stiffer parts of the brush
substrate.
It is clear that this `force' becomes significantly
smaller when the droplet enters the stiffer half of the 
substrate in the $x$ direction. However, this
small `force' is still able to sustain the 
motion of the droplet and lead to a successful
durotaxis case. Still, the
inset of Fig.~\ref{fig:6}b reflects
the observed behavior caused by an
average `force' that propels the droplet.
Finally, we have clearly seen
from our data that unsuccessful durotaxis cases
are characterised by a flat interfacial energy
that fluctuates around a constant value, which
would ideally yield $f_{\rm x} \approx 0$.

To further understand the durotaxis mechanism of
droplets on brush substrates, we have gone one
step further and tried to identify the origin
of the changing interfacial energy in successful
durotaxis cases.
In particular, we have measured the standard deviation
of the end-to-end distance, $\Omega$, 
which describes the width of the free-end positions 
distribution
of the brush polymer chains (Fig.~\ref{fig:7}a), i.e., 
the surface roughness of the brush.
We can observe that $\Omega$
decays monotonically towards the stiffest parts of
the brush. In other words, stiffer polymer chains
exhibit a smaller extent of fluctuations concerning the 
end-to-end distance of the polymer chains, which
is generally expected when the stiffness increases. 
Moreover, we can see that the decay of $\Omega$
is faster at the soft parts of the substrate and
generally follows the decay in the interfacial
energy (Fig.~\ref{fig:6}). 
The inset of Fig.~\ref{fig:7}a presents results
for the the density profile
in the $z$ direction at different positions $X$ in
the $x$ direction (along the stiffness gradient).
We observe that the thickness of the brush surface becomes
smaller towards the stiffer parts,
which would correspond to a flatter surface locally. 
In contrast, slower decaying density profiles
correspond to a larger thickness of the brush surface,
in other words, to a rougher brush surface,
which is observed in the soft substrate parts.
In practice, rough surfaces result in a smaller
number of contacts with the droplet, and as
a result a higher (less negative) interfacial
energy.\cite{Theodorakis2017}
On the contrary, a flat profile would allow
for a larger number of contacts  between the droplet and the
substrate. For this reason, the droplet moves
towards the stiffer parts of the substrate. 
We provide further evidence for our argument by measuring the 
interpenetration length, $W$, as a function of the
position, $X$, of the centre-of-mass of the droplet
along the substrate in the direction 
of the stiffness gradient (Fig.~\ref{fig:7}b).
This property reflects
the average distance of the substrate--droplet
contact pairs, which is noted here with the symbol
$l_z$. We can clearly see that the 
interpenetration length decreases at the
stiffer parts of the substrate, which points to
a sharper (flatter) brush surface, in accordance with the
results of Fig.~\ref{fig:7}a.
We have further explored the dependence of $W$
on $\varepsilon_{\rm db}$ 
and $\sigma_g$ and
found that it decreases as a function of $\sigma_{\rm g}$
due to the induced stiffness by the steric interactions
between the brush polymer chains, while it increases
with $\varepsilon_{\rm db}$. An important conclusion 
from the results of Fig.~\ref{fig:7}c is that $W$ is
clearly larger in the case of soft brushes 
($k_{\theta}=0~\epsilon/{\rm rad}^2$). However, 
it is the gradual change of this roughness that
plays an important role in inducing the interfacial
gradient, which in turn translates into the 
effective `force' that drives the droplet motion.

\begin{figure}[bt!]
\centering
 \includegraphics[width=\columnwidth]{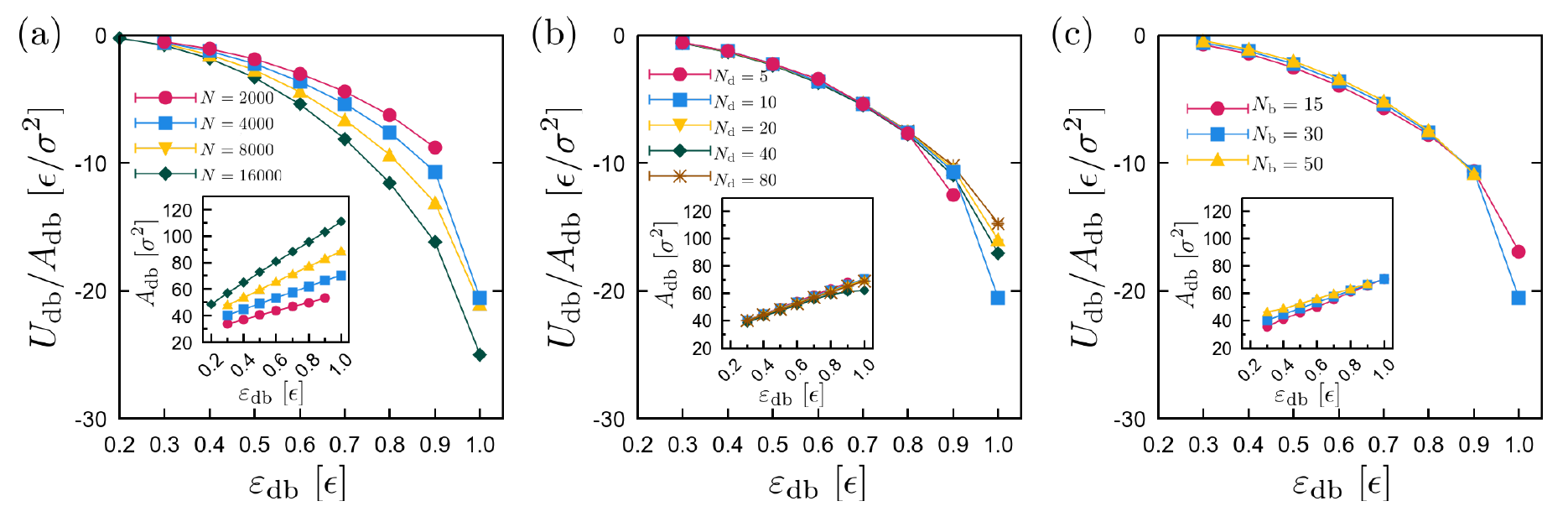}
\caption{\label{fig:8} Droplet--brush interfacial energy,
$U_{\rm db}$, per area of the droplet--substrate contact surface,
$A_{\rm db}$,
as a function of $\varepsilon_{\rm db}$. 
$\sigma_{\rm g} = 0.6~\sigma^{-2}$ and 
$\Gamma = 0.8~\epsilon/{\rm rad}^2\sigma$.
Panels show results for different (a) $N$
($N_{\rm b}=30$ and $N_{\rm d}=10$ beads),
(b) $N_{\rm d}$ ($N_{\rm b}=30$ beads), 
and (c) $N_{\rm b}$ ($N_{\rm d}=10$ beads).
The area, $A_{\rm db}$, was
calculated by using the qhull library\cite{qhull1996}, 
taking the area of the convex hull.
}
\end{figure}

\begin{figure}[bt!]
\centering
 \includegraphics[width=0.9\columnwidth]{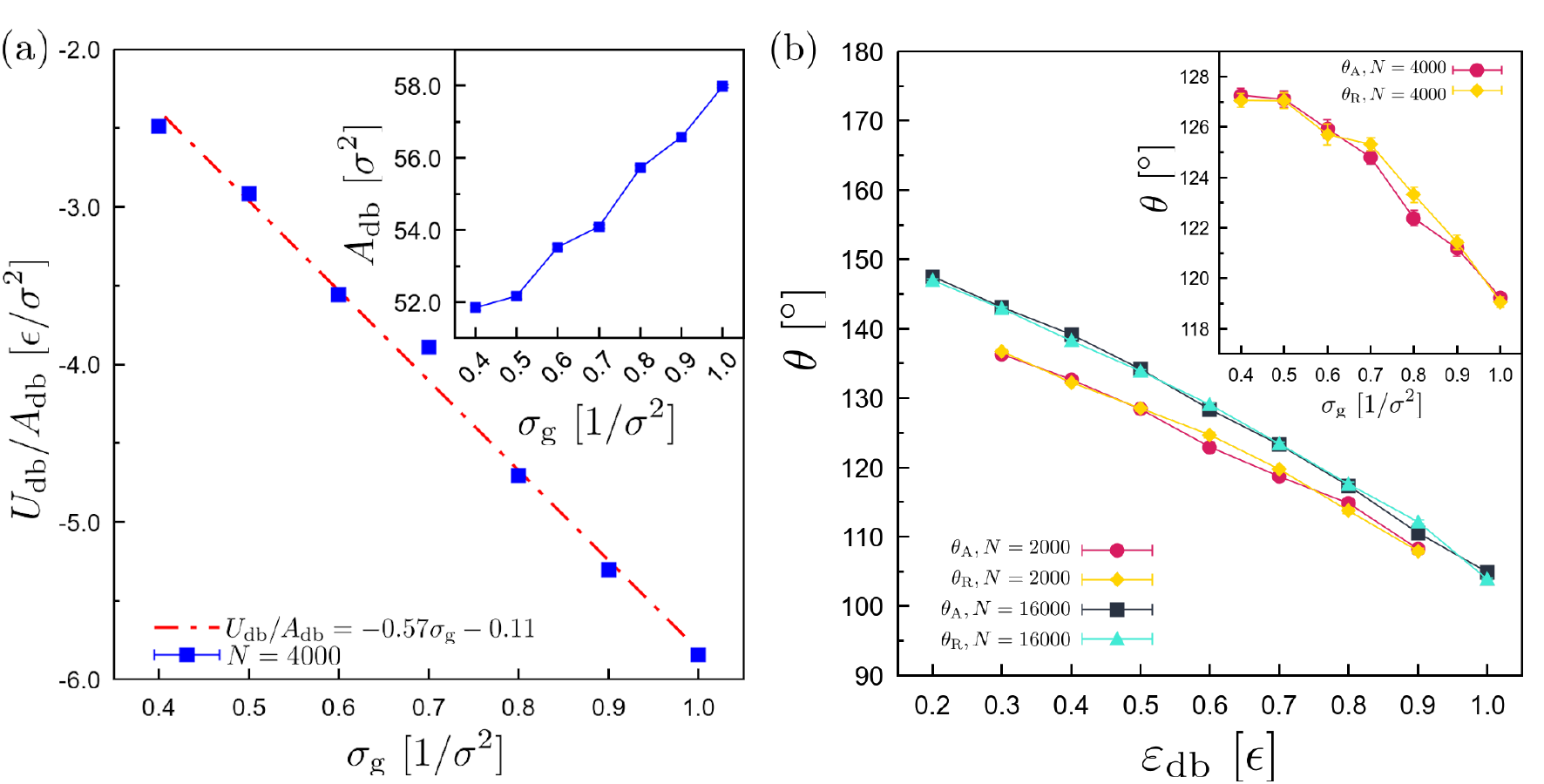}
\caption{\label{fig:9} (a) Droplet--brush average
interfacial energy,
$U_{\rm db}$, per area, $A_{\rm db}$, 
as a function of the grafting density, $\sigma_{\rm g}$.
Dashed-dotted line is the result of a linear fit
with slope $-0.57~\epsilon$ as indicated.
Inset illustrates the values of the area, $A_{\rm db}$,
versus the grafting density.
$\varepsilon_{\rm db} = 0.6~\epsilon$, $N=4000$,
$N_{\rm b}=30$ and $N_{\rm d}=10$ beads. 
$\Gamma = 0.8~\epsilon/{\rm rad}^2\sigma$.
(b) Advancing, $\theta_{\rm A}$, and
receding, $\theta_{\rm R}$, contact
angles along the durotaxis motion ($x$ direction)
as a function of $\varepsilon_{\rm db}$ for droplets
of different size as indicated. 
$\sigma_{\rm g}=0.6~\sigma^{-2}$. 
The inset shows the
dependence of the contact angles on the 
grafting density, $\sigma_{\rm g}$, for 
$\varepsilon_{\rm db}=0.6~\epsilon$. 
In both the main panel and the inset,
$N_{\rm d}=10$ and $N_{\rm b}=30$ beads,
and $\Gamma = 0.8~\epsilon/{\rm rad}^2\sigma$.
}
\end{figure}

With these indications that
the minimization of the interfacial energy
drives the droplet toward areas of more negative
energy, in Fig.~\ref{fig:8}, we show the interfacial
energy for the same systems shown in Fig.~\ref{fig:5}.
When the chain length of the droplet polymer chains,
$N_{\rm d}$, varies within the range considered in
our study, there are no noticeable changes
in the interfacial energy (Fig.~\ref{fig:8}b)
and systems are rather characterised by
an equivalent average energy profile,
which is reflected in the average velocity of the
droplets (Fig.~\ref{fig:5}b). There, we have found
that a change in the viscosity of the droplet does
not significantly affect the durotaxis efficiency.
Slight changes in the average interfacial energy are
observed when the brush chain length, $N_{\rm b}$,
varies (Fig.~\ref{fig:8}c). 
These results are also consistent with those
of Fig.~\ref{fig:5}c. In particular, we can see
that the brush with chain length $N_{\rm b}=50$ beads
appears to have the lowest average interfacial energy and
the largest interfacial area for the whole range of
$\varepsilon_{\rm db}$ considered here. 
Moreover, differences in the interfacial energy between
the systems with different $N_{\rm b}$ become smaller
as $\varepsilon_{\rm db}$ increases, which is
much clearer in the data concerning the
interfacial area, $A_{\rm db}$, in line
with the results of Fig.~\ref{fig:5}c.
In contrast, rather larger differences are observed when the
size of the droplets, $N$, changes (Fig.~\ref{fig:8}a).
In particular, droplets of smaller size have
a markedly lower (less negative)
interfacial energy per area than larger
droplets. These differences become
more apparent as the strength of interactions,
$\varepsilon_{\rm db}$, increases. 
Interestingly, although the energy here shows a monotonic
behavior, this is not the case for the average
velocity of the droplet (Fig.~\ref{fig:5}a),
which indicates that droplets of different size
are differently affected in a global sense 
for a given set of substrate parameters. 
One contributing factor is that the change
in interfacial energy from soft to hard substrate
is proportional to the contact area $\propto N^{2/3}$, 
while the mass is proportional to $N$. 
Therefore the energy change per atom is $\propto N^{-1/3}$, 
other factors being equal, and the velocity 
change can be expected to scale as $N^{-1/6}$. 
This is not inconsistent with Fig.~\ref{fig:5}a
at medium strength $\varepsilon_{\rm db}$.
Finally, the interfacial energy decreases proportionally
to the grafting density, with a slope 
of $-0.57~\epsilon$ (Fig.~\ref{fig:9}a),
suggesting a proportionally larger number of contacts
between the droplet and the substrate as the
grafting density grows.

We have also monitored the advancing, $\theta_{\rm A}$, and
receding, $\theta_{\rm R}$, contact angles during the
durotaxial motion and results are presented in
Fig.~\ref{fig:9}b for typical cases. 
In this case, the angles
have been determined by using the curvature of the
droplet as described in previous 
studies\cite{Theodorakis2015},
thus avoiding error-prone fits.
We observe that both $\theta_{\rm A}$ and $\theta_{\rm R}$
decrease rather linearly
with the increase of $\varepsilon_{\rm db}$.
A linear decrease has been observed for droplets on solid 
substrates.\cite{Theodorakis2015,Theodorakis2017}
Droplets of different size, namely $N=2000$ and
$N=16000$ beads show the same trend. Moreover, as
$\varepsilon_{\rm db}$ increases, the differences
between smaller and larger droplets become
less pronounced.
A much weaker dependence of the contact angles
on the grafting density has been observed for
$\sigma_{\rm g} \leq 0.6~\sigma^{-2}$
and a small scale rather linear dependence for 
$\sigma_{\rm g} \geq 0.6~\sigma^{-2}$. 
Our results, which are averaged over the whole 
trajectory, do not show any statistically 
significant difference between advancing
and receding contact angles. 
Due to the greater magnitude of interfacial energy 
(more negative) at the stiffer parts of the substrate
(Fig.~\ref{fig:6}) one would expect 
a smaller advancing contact
angle (for example, as Fig.~\ref{fig:9}b suggests,
larger attraction leads to smaller contact angles). 
Slightly larger values for the 
receding contact angle, $\theta_{\rm R}$, are observed
consistently for different values of $\sigma_{\rm g}$
and $\varepsilon_{\rm db}$, but, still,
within the statistical error. 
In experiment and theory, when
the droplet moves by steadily applied external 
force the receding contact angles are smaller
than the advancing ones in which case
friction effects might also play a role.

\begin{figure}[bt!]
\centering
 \includegraphics[width=1.0\columnwidth]{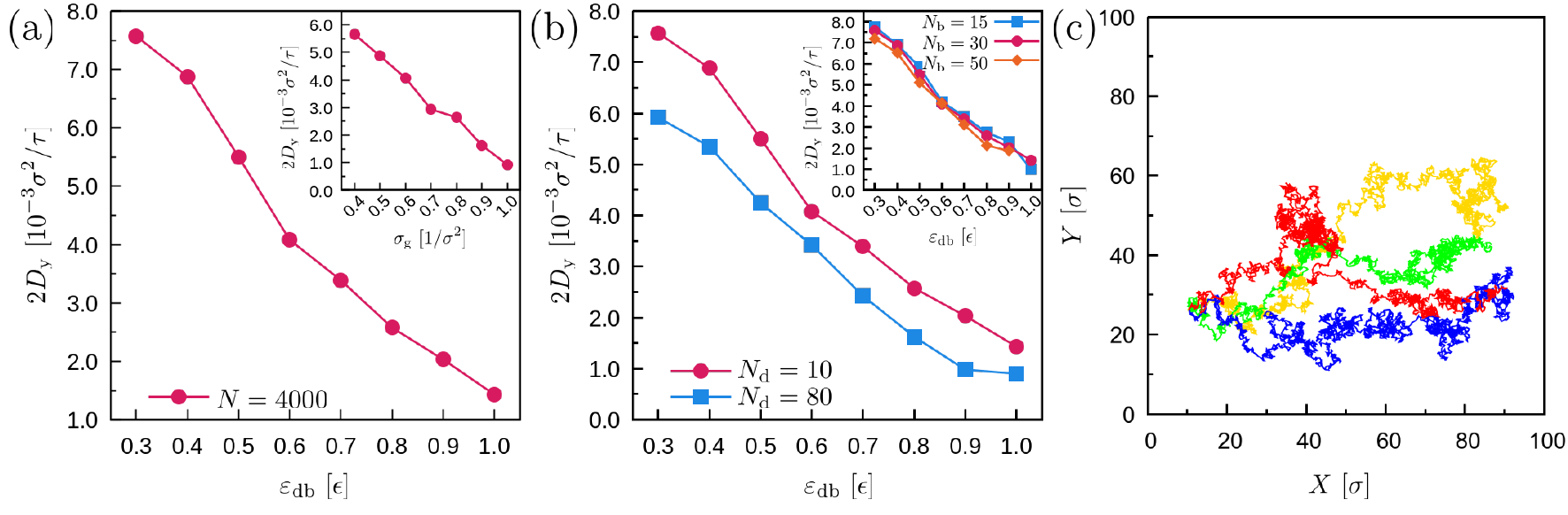}
\caption{\label{fig:10} Diffusion coefficient 
($\sigma_{\rm g}=0.6~\sigma^{-2}$) as a function
of (a) $\varepsilon_{\rm db}$ (the inset shows the dependence
on the grafting density, $\sigma_{\rm g}$ for the case
$\varepsilon_{\rm db}=0.6~\epsilon$). 
$N_{\rm d}=10$ and $N_{\rm b}=30$ beads.
(b) Diffusion coefficient for different
droplet chain length, $N_{\rm d}$,
brush chain length, $N_{\rm b}$ (inset)
as a function of $\varepsilon_{\rm db}$.
$\sigma_{\rm g}=0.6~\sigma^{-2}$, 
$\varepsilon_{\rm db}=0.6~\epsilon$ in the inset.
$N=4000$ beads.
(c) Typical, successful durotaxis trajectories of
the centre of mass of the droplet in the $x-y$ space 
as indicated by different colours for separate run.
$\sigma_{\rm g}=0.6~\sigma^{-2}$,
$\varepsilon_{\rm db} = 0.6~\epsilon$, $N=4000$,
$N_{\rm b}=30$ and $N_{\rm d}=10$ beads, and
$\Gamma = 0.8~\epsilon/{\rm rad}^2\sigma$.
}
\end{figure}

Finally, to investigate the diffusion of the droplet, 
and how it is affected by the main parameters, 
we have analysed the movement
of the centre of mass of the droplet in the $y$ direction
(Fig.~\ref{fig:10}). 
Dynamics in the $y$ direction is free 
of extraneous effects due to grafting density gradient
and the durotaxis itself,
so serves as a good control and
measurement of the diffusion properties.
Overall droplets with higher viscosity and
larger wettability on the brush substrate, show
a slower diffusion. The same effect has the
grafting density. In contrast, $N_{\rm b}$ 
does not have a tangible effect on the droplet
motion. In Fig.~\ref{fig:10}c, we present 
typical trajectories as a function of the $X$
and $Y$ positions of the droplet centre-of-mass. 
We can observe that durotaxis motion is also characterised
by significant random motion in both $x$ and $y$ axes 
including `reversals'. 
In some cases, we can observe that the droplet
can cover in the $y$ direction
a distance as much as 30\%  of the
total distance in the $x$ direction, which 
suggests that durotaxis motion 
is also strongly affected by random effects at the interface
between the droplet and the substrate,
especially when the gradient of the interfacial energy
becomes small, \textit{i.e}, the driving force of 
the durotaxis is correspondingly small.



\section{CONCLUSIONS}

In this study, we have proposed a new design 
of brush polymer substrates that
is capable of leading to the durotaxis motion of nanodroplets.
The knowledge gained here may lead to new experimental
brush-based substrate designs 
and provide further understanding
of relevant biological processes, such as the
motion of cells on tissues\cite{DuChez2019,Khang2015},
or the mucus flow around lung cilia\cite{Button2012}.
Our analysis has also indicated that the durotaxis 
motion on brush substrates is driven by a corresponding
gradient in the interfacial energy between the droplet
and the substrate, in line with previous findings 
in the context of various other substrate 
designs.\cite{Theodorakis2017,Theodorakis2022,Chang2015}
Moreover, we have found that the origin 
of the steady increase of the interfacial 
energy is related to the state of 
the brush surface,
which appears as a `rougher' profile in the softer parts
of the substrate and a flatter interface in the
stiffer regions. This translates into a larger number
of contacts between the droplet and the substrate
in the stiffer parts, and hence a more negative
interfacial energy along the direction of the
stiffness gradient. 

We have also conducted a parametric study based
on the various system parameters in order to
gain further insight into the system and identify
the key parameters of the brush substrate design. 
Two of the key
parameters are the grafting density of the polymer
chains and the substrate wettability. 
Our findings suggest that the durotaxis motion
is favored by a moderate grafting density, which
in our case translates to values of
grafting density $\sigma_{\rm g} \approx 0.5~\sigma^{-2}$,
and the same time moderate values 
of the substrate wettability,
namely $0.5~\epsilon \leq \varepsilon_{\rm db} \leq 0.9~\epsilon$.
Surprisingly, we have found that the stiffness
gradient itself, as defined by the linear change
in the stiffness of individual polymer chains 
tuned by the harmonic constant, $k_{\rm \theta}$,
does not induce important changes in the 
efficiency of the durotaxis motion. 
This might be due to the relatively large
grafting density, which affects the apparent
stiffness of the polymer chains, thus minimising 
the actual effect of the gradient. Such an effect
has not been observed in previous systems.\cite{Theodorakis2017}
Moreover, the droplet viscosity also seems not to
affect the durotaxis efficiency as also the length of the polymer
brush chains, since the durotaxis performance seems
to saturate after a certain polymer length. 
In contrast, the size of the droplet plays a
role. In particular, smaller droplets seem to reach
a faster durotaxis and for a lower adhesion to the
subtrate, while larger droplets move with
a lower average velocity exhibiting their maximum
velocity at larger droplet--substrate adhesion. 
Furhtermore, we have not identified any tangible
differences between the advancing and receding
contact angles during the durotaxis motion. 
The durotaxis motion is induced by tiny effects
at the droplet--substrate interface as 
judged by the gradient in the droplet--substrate
interfacial energy. Hence,
there is a back and forth wiggling motion while
the brush slowly guides the droplet towards 
a lower energy state. Finally, we have
discussed how our findings could motivate
further experimental research in the area
of self-sustained fluid motion on brush
gradient substrates. 
It would also be interesting to investigate
the droplet durotaxis behavior when many droplets
are placed onto the substrate and explore various
effects, such as droplet coalescence or how the 
droplet--substrate interactions are affected
in populations of droplets. 
Further work
to explore such possibilities is
expected in this direction in the future.
We anticipate that this study for the first time presents
new possibilities for the implementation
and understanding of durotaxis motion of fluids
on brush substrates with important implications for
various areas, for example, in the context of biology.

\begin{acknowledgement}

This research has been supported by the National Science Centre, Poland, under grant
No. 2019/35/B/ST3/03426. A. M. acknowledges support by COST (European Cooperation in
Science and Technology [See http://www.cost.eu and https://www.fni.bg] and its 
Bulgarian partner FNI/MON under KOST-11). 
This research was supported in part by PLGrid Infrastructure.

\end{acknowledgement}

\begin{suppinfo}

M*.mp4: Movie illustrates an example of
durotaxis motion onto a gradient brush substrate. 
The parameters for the system are: $N=4000$, $N_{\rm d}=10$,
and $N_{\rm b}=30$ beads. $\sigma_{\rm g}=0.6~\sigma^{-2}$, 
$\Gamma = 0.8~\epsilon/{\rm rad}^2 \sigma$ (chains at the 
very soft right end are fully flexible, \textit{i.e.}, 
$k_{\theta}=0~\epsilon/{\rm rad}^2$, while at the very 
stiff part $k_{\theta}=80~\epsilon/{\rm rad}^2$), 
and $\varepsilon_{\rm db}=0.6~\epsilon$.
\end{suppinfo}


\providecommand{\latin}[1]{#1}
\makeatletter
\providecommand{\doi}
  {\begingroup\let\do\@makeother\dospecials
  \catcode`\{=1 \catcode`\}=2 \doi@aux}
\providecommand{\doi@aux}[1]{\endgroup\texttt{#1}}
\makeatother
\providecommand*\mcitethebibliography{\thebibliography}
\csname @ifundefined\endcsname{endmcitethebibliography}
  {\let\endmcitethebibliography\endthebibliography}{}


\providecommand{\latin}[1]{#1}
\makeatletter
\providecommand{\doi}
  {\begingroup\let\do\@makeother\dospecials
  \catcode`\{=1 \catcode`\}=2 \doi@aux}
\providecommand{\doi@aux}[1]{\endgroup\texttt{#1}}
\makeatother
\providecommand*\mcitethebibliography{\thebibliography}
\csname @ifundefined\endcsname{endmcitethebibliography}
  {\let\endmcitethebibliography\endthebibliography}{}
\begin{mcitethebibliography}{66}
\providecommand*\natexlab[1]{#1}
\providecommand*\mciteSetBstSublistMode[1]{}
\providecommand*\mciteSetBstMaxWidthForm[2]{}
\providecommand*\mciteBstWouldAddEndPuncttrue
  {\def\EndOfBibitem{\unskip.}}
\providecommand*\mciteBstWouldAddEndPunctfalse
  {\let\EndOfBibitem\relax}
\providecommand*\mciteSetBstMidEndSepPunct[3]{}
\providecommand*\mciteSetBstSublistLabelBeginEnd[3]{}
\providecommand*\EndOfBibitem{}
\mciteSetBstSublistMode{f}
\mciteSetBstMaxWidthForm{subitem}{(\alph{mcitesubitemcount})}
\mciteSetBstSublistLabelBeginEnd
  {\mcitemaxwidthsubitemform\space}
  {\relax}
  {\relax}

\bibitem[Srinivasarao \latin{et~al.}(2001)Srinivasarao, Collings, Philips, and
  Patel]{Srinivasarao2001}
Srinivasarao,~M.; Collings,~D.; Philips,~A.; Patel,~S. Three-dimensionally
  ordered array of air bubbles in a polymer film. \emph{Science} \textbf{2001},
  \emph{292}, 79--83\relax
\mciteBstWouldAddEndPuncttrue
\mciteSetBstMidEndSepPunct{\mcitedefaultmidpunct}
{\mcitedefaultendpunct}{\mcitedefaultseppunct}\relax
\EndOfBibitem
\bibitem[Chaudhury and Whitesides(1992)Chaudhury, and
  Whitesides]{Chaudhury1992}
Chaudhury,~M.~K.; Whitesides,~G.~M. How to Make Water Run Uphill.
  \emph{Science} \textbf{1992}, \emph{256}, 1539--1541\relax
\mciteBstWouldAddEndPuncttrue
\mciteSetBstMidEndSepPunct{\mcitedefaultmidpunct}
{\mcitedefaultendpunct}{\mcitedefaultseppunct}\relax
\EndOfBibitem
\bibitem[Wong \latin{et~al.}(2011)Wong, Kang, Tang, Smythe, Hatton, Grinthal,
  and Aizenberg]{Wong2011}
Wong,~T.-S.; Kang,~S.~H.; Tang,~S. K.~Y.; Smythe,~E.~J.; Hatton,~B.~D.;
  Grinthal,~A.; Aizenberg,~J. Bioinspired self-repairing slippery surfaces with
  pressure-stable omniphobicity. \emph{Nature} \textbf{2011}, \emph{477},
  443--447\relax
\mciteBstWouldAddEndPuncttrue
\mciteSetBstMidEndSepPunct{\mcitedefaultmidpunct}
{\mcitedefaultendpunct}{\mcitedefaultseppunct}\relax
\EndOfBibitem
\bibitem[Lagubeau \latin{et~al.}(2011)Lagubeau, Le~Merrer, Clanet, and
  Qu\'er\'e]{Lagubeau2011}
Lagubeau,~G.; Le~Merrer,~M.; Clanet,~C.; Qu\'er\'e,~D. Leidenfrost on a
  ratchet. \emph{Nat. Phys.} \textbf{2011}, \emph{7}, 395--398\relax
\mciteBstWouldAddEndPuncttrue
\mciteSetBstMidEndSepPunct{\mcitedefaultmidpunct}
{\mcitedefaultendpunct}{\mcitedefaultseppunct}\relax
\EndOfBibitem
\bibitem[Prakash \latin{et~al.}(2008)Prakash, Qu\'er\'e, and Bush]{Prakash2008}
Prakash,~M.; Qu\'er\'e,~D.; Bush,~J.~W. Surface tension transport of prey by
  feeding shorebirds: The capillary ratchet. \emph{Science} \textbf{2008},
  \emph{320}, 931--934\relax
\mciteBstWouldAddEndPuncttrue
\mciteSetBstMidEndSepPunct{\mcitedefaultmidpunct}
{\mcitedefaultendpunct}{\mcitedefaultseppunct}\relax
\EndOfBibitem
\bibitem[Darhuber and Troian(2005)Darhuber, and Troian]{Darhuber2005}
Darhuber,~A.; Troian,~S. Principles of microfluidic actuation by modulation of
  surface stresses. \emph{Annu. Rev. Fluid Mech.} \textbf{2005}, \emph{37},
  425--455\relax
\mciteBstWouldAddEndPuncttrue
\mciteSetBstMidEndSepPunct{\mcitedefaultmidpunct}
{\mcitedefaultendpunct}{\mcitedefaultseppunct}\relax
\EndOfBibitem
\bibitem[Yao and Bowick(2012)Yao, and Bowick]{Yao2012}
Yao,~Z.; Bowick,~M.~J. {Self-propulsion of droplets by spatially-varying
  surface topography}. \emph{Soft Matter} \textbf{2012}, \emph{8},
  1142--1145\relax
\mciteBstWouldAddEndPuncttrue
\mciteSetBstMidEndSepPunct{\mcitedefaultmidpunct}
{\mcitedefaultendpunct}{\mcitedefaultseppunct}\relax
\EndOfBibitem
\bibitem[Li \latin{et~al.}(2018)Li, Yan, Fichthorn, and Yu]{Li2018}
Li,~H.; Yan,~T.; Fichthorn,~K.~A.; Yu,~S. {Dynamic contact angles and
  mechanisms of motion of water droplets moving on nano-pillared
  superhydrophobic surfaces: A molecular dynamics simulation study}.
  \emph{Langmuir} \textbf{2018}, \emph{34}, 9917--9926\relax
\mciteBstWouldAddEndPuncttrue
\mciteSetBstMidEndSepPunct{\mcitedefaultmidpunct}
{\mcitedefaultendpunct}{\mcitedefaultseppunct}\relax
\EndOfBibitem
\bibitem[Becton and Wang(2016)Becton, and Wang]{Becton2016}
Becton,~M.; Wang,~X. {Controlling nanoflake motion using stiffness gradients on
  hexagonal boron nitride}. \emph{RSC Adv.} \textbf{2016}, \emph{6},
  51205--51210\relax
\mciteBstWouldAddEndPuncttrue
\mciteSetBstMidEndSepPunct{\mcitedefaultmidpunct}
{\mcitedefaultendpunct}{\mcitedefaultseppunct}\relax
\EndOfBibitem
\bibitem[van~den Heuvel and Dekker(2007)van~den Heuvel, and
  Dekker]{vandenHeuvel2007}
van~den Heuvel,~M. G.~L.; Dekker,~C. Motor proteins at work for nanotechnology.
  \emph{Science} \textbf{2007}, \emph{317}, 333--336\relax
\mciteBstWouldAddEndPuncttrue
\mciteSetBstMidEndSepPunct{\mcitedefaultmidpunct}
{\mcitedefaultendpunct}{\mcitedefaultseppunct}\relax
\EndOfBibitem
\bibitem[DuChez \latin{et~al.}(2019)DuChez, Doyle, Dimitriadis, and
  Yamada]{DuChez2019}
DuChez,~B.~J.; Doyle,~A.~D.; Dimitriadis,~E.~K.; Yamada,~K.~M. Durotaxis by
  human cancer cells. \emph{Biophys. J.} \textbf{2019}, \emph{116},
  670--683\relax
\mciteBstWouldAddEndPuncttrue
\mciteSetBstMidEndSepPunct{\mcitedefaultmidpunct}
{\mcitedefaultendpunct}{\mcitedefaultseppunct}\relax
\EndOfBibitem
\bibitem[Khang(2015)]{Khang2015}
Khang,~G. Evolution of gradient concept for the application of regenerative
  medicine. \emph{Biosurface Biotribology} \textbf{2015}, \emph{1},
  202--213\relax
\mciteBstWouldAddEndPuncttrue
\mciteSetBstMidEndSepPunct{\mcitedefaultmidpunct}
{\mcitedefaultendpunct}{\mcitedefaultseppunct}\relax
\EndOfBibitem
\bibitem[Theodorakis \latin{et~al.}(2017)Theodorakis, Egorov, and
  Milchev]{Theodorakis2017}
Theodorakis,~P.~E.; Egorov,~S.~A.; Milchev,~A. Stiffness-guided motion of a
  droplet on a solid substrate. \emph{J. Chem. Phys.} \textbf{2017},
  \emph{146}, 244705\relax
\mciteBstWouldAddEndPuncttrue
\mciteSetBstMidEndSepPunct{\mcitedefaultmidpunct}
{\mcitedefaultendpunct}{\mcitedefaultseppunct}\relax
\EndOfBibitem
\bibitem[Lo \latin{et~al.}(2000)Lo, Wang, Dembo, and Wang]{Lo2000}
Lo,~C.-M.; Wang,~H.-B.; Dembo,~M.; Wang,~Y.-L. Cell movement is guided by the
  rigidity of the substrate. \emph{Biophys. J.} \textbf{2000}, \emph{79},
  144--152\relax
\mciteBstWouldAddEndPuncttrue
\mciteSetBstMidEndSepPunct{\mcitedefaultmidpunct}
{\mcitedefaultendpunct}{\mcitedefaultseppunct}\relax
\EndOfBibitem
\bibitem[Style \latin{et~al.}(2013)Style, Che, Park, Weon, Je, Hyland, German,
  Power, Wilen, Wettlaufer, and Dufresne]{Style2013}
Style,~R.~W.; Che,~Y.; Park,~S.~J.; Weon,~B.~M.; Je,~J.~H.; Hyland,~C.;
  German,~G.~K.; Power,~M.~P.; Wilen,~L.~A.; Wettlaufer,~J.~S.; Dufresne,~E.~R.
  Patterning droplets with durotaxis. \emph{Proc. Natl. Acad. Sci. U.S.A.}
  \textbf{2013}, \emph{110}, 12541--12544\relax
\mciteBstWouldAddEndPuncttrue
\mciteSetBstMidEndSepPunct{\mcitedefaultmidpunct}
{\mcitedefaultendpunct}{\mcitedefaultseppunct}\relax
\EndOfBibitem
\bibitem[Chang \latin{et~al.}(2015)Chang, Zhang, Guo, Guo, and Gao]{Chang2015}
Chang,~T.; Zhang,~H.; Guo,~Z.; Guo,~X.; Gao,~H. {Nanoscale directional motion
  towards regions of stiffness}. \emph{Phys. Rev. Lett.} \textbf{2015},
  \emph{114}, 015504\relax
\mciteBstWouldAddEndPuncttrue
\mciteSetBstMidEndSepPunct{\mcitedefaultmidpunct}
{\mcitedefaultendpunct}{\mcitedefaultseppunct}\relax
\EndOfBibitem
\bibitem[Pham \latin{et~al.}(2016)Pham, Xue, {Del Campo}, and
  Salierno]{Pham2016}
Pham,~J.~T.; Xue,~L.; {Del Campo},~A.; Salierno,~M. {Guiding cell migration
  with microscale stiffness patterns and undulated surfaces}. \emph{Acta
  Biomaterialia} \textbf{2016}, \emph{38}, 106--115\relax
\mciteBstWouldAddEndPuncttrue
\mciteSetBstMidEndSepPunct{\mcitedefaultmidpunct}
{\mcitedefaultendpunct}{\mcitedefaultseppunct}\relax
\EndOfBibitem
\bibitem[Lazopoulos and Stamenovi{\'{c}}(2008)Lazopoulos, and
  Stamenovi{\'{c}}]{Lazopoulos2008}
Lazopoulos,~K.~A.; Stamenovi{\'{c}},~D. {Durotaxis as an elastic stability
  phenomenon}. \emph{J. Biomech.} \textbf{2008}, \emph{41}, 1289--1294\relax
\mciteBstWouldAddEndPuncttrue
\mciteSetBstMidEndSepPunct{\mcitedefaultmidpunct}
{\mcitedefaultendpunct}{\mcitedefaultseppunct}\relax
\EndOfBibitem
\bibitem[Becton and Wang(2014)Becton, and Wang]{Becton2014}
Becton,~M.; Wang,~X. {Thermal gradients on graphene to drive nanoflake motion}.
  \emph{J. Chem. Theory Comput.} \textbf{2014}, \emph{10}, 722--730\relax
\mciteBstWouldAddEndPuncttrue
\mciteSetBstMidEndSepPunct{\mcitedefaultmidpunct}
{\mcitedefaultendpunct}{\mcitedefaultseppunct}\relax
\EndOfBibitem
\bibitem[Barnard(2015)]{Barnard2015}
Barnard,~A.~S. {Nanoscale locomotion without fuel}. \emph{Nature}
  \textbf{2015}, \emph{519}, 37--38\relax
\mciteBstWouldAddEndPuncttrue
\mciteSetBstMidEndSepPunct{\mcitedefaultmidpunct}
{\mcitedefaultendpunct}{\mcitedefaultseppunct}\relax
\EndOfBibitem
\bibitem[Palaia \latin{et~al.}(2021)Palaia, Paraschiv, Debets, Storm, and
  \v{S}ari\'c]{Palaia2021}
Palaia,~I.; Paraschiv,~A.; Debets,~V.~E.; Storm,~C.; \v{S}ari\'c,~A. Durotaxis
  of Passive Nanoparticles on Elastic Membranes. \emph{ACS Nano} \textbf{2021},
  \emph{15}, 15794--15802\relax
\mciteBstWouldAddEndPuncttrue
\mciteSetBstMidEndSepPunct{\mcitedefaultmidpunct}
{\mcitedefaultendpunct}{\mcitedefaultseppunct}\relax
\EndOfBibitem
\bibitem[Tamim and Bostwick(2021)Tamim, and Bostwick]{Tamim2021}
Tamim,~S.~I.; Bostwick,~J.~B. Model of spontaneous droplet transport on a soft
  viscoelastic substrate with nonuniform thickness. \emph{Phys. Rev. E}
  \textbf{2021}, \emph{104}, 034611\relax
\mciteBstWouldAddEndPuncttrue
\mciteSetBstMidEndSepPunct{\mcitedefaultmidpunct}
{\mcitedefaultendpunct}{\mcitedefaultseppunct}\relax
\EndOfBibitem
\bibitem[Bardall \latin{et~al.}(2020)Bardall, Chen, Daniels, and
  Shearer]{Bardall2020}
Bardall,~A.; Chen,~S.-Y.; Daniels,~K.~E.; Shearer,~M. {Gradient-induced droplet
  motion over soft solids}. \emph{IMA J. Appl. Math} \textbf{2020}, \emph{85},
  495--512\relax
\mciteBstWouldAddEndPuncttrue
\mciteSetBstMidEndSepPunct{\mcitedefaultmidpunct}
{\mcitedefaultendpunct}{\mcitedefaultseppunct}\relax
\EndOfBibitem
\bibitem[Theodorakis \latin{et~al.}(2002)Theodorakis, Egorov, and
  Milchev]{Theodorakis2022}
Theodorakis,~P.~E.; Egorov,~S.~A.; Milchev,~A. Rugotaxis: Droplet motion
  without external energy supply. \emph{EPL} \textbf{2002}, \emph{137},
  43002\relax
\mciteBstWouldAddEndPuncttrue
\mciteSetBstMidEndSepPunct{\mcitedefaultmidpunct}
{\mcitedefaultendpunct}{\mcitedefaultseppunct}\relax
\EndOfBibitem
\bibitem[Hiltl and B\"oker(2016)Hiltl, and B\"oker]{Hiltl2016}
Hiltl,~S.; B\"oker,~A. {Wetting Phenomena on (Gradient) Wrinkle Substrates}.
  \emph{Langmuir} \textbf{2016}, \emph{32}, 8882--8888\relax
\mciteBstWouldAddEndPuncttrue
\mciteSetBstMidEndSepPunct{\mcitedefaultmidpunct}
{\mcitedefaultendpunct}{\mcitedefaultseppunct}\relax
\EndOfBibitem
\bibitem[Pismen and Thiele(2006)Pismen, and Thiele]{Pismen2006}
Pismen,~L.~M.; Thiele,~U. {Asymptotic theory for a moving droplet driven by a
  wettability gradient}. \emph{Phys. Fluids} \textbf{2006}, \emph{18},
  042104\relax
\mciteBstWouldAddEndPuncttrue
\mciteSetBstMidEndSepPunct{\mcitedefaultmidpunct}
{\mcitedefaultendpunct}{\mcitedefaultseppunct}\relax
\EndOfBibitem
\bibitem[Wu \latin{et~al.}(2017)Wu, Zhu, Cao, Zhang, Wu, Liang, Chai, and
  Liu]{Wu2017}
Wu,~H.; Zhu,~K.; Cao,~B.; Zhang,~Z.; Wu,~B.; Liang,~L.; Chai,~G.; Liu,~A. Smart
  design of wettability-patterned gradients on substrate-independent coated
  surfaces to control unidirectional spreading of droplets. \emph{Soft Matter}
  \textbf{2017}, \emph{13}, 2995--3002\relax
\mciteBstWouldAddEndPuncttrue
\mciteSetBstMidEndSepPunct{\mcitedefaultmidpunct}
{\mcitedefaultendpunct}{\mcitedefaultseppunct}\relax
\EndOfBibitem
\bibitem[Theodorakis \latin{et~al.}(2021)Theodorakis, Amirfazli, Hu, and
  Che]{Theodorakis2021}
Theodorakis,~P.~E.; Amirfazli,~A.; Hu,~B.; Che,~Z. Droplet Control Based on
  Pinning and Substrate Wettability. \emph{Langmuir} \textbf{2021}, \emph{37},
  4248--4255\relax
\mciteBstWouldAddEndPuncttrue
\mciteSetBstMidEndSepPunct{\mcitedefaultmidpunct}
{\mcitedefaultendpunct}{\mcitedefaultseppunct}\relax
\EndOfBibitem
\bibitem[Feng \latin{et~al.}(2020)Feng, Delannoy, Malod, Zheng, Qu\'er\'e, and
  Wang]{Feng2020}
Feng,~S.; Delannoy,~J.; Malod,~A.; Zheng,~H.; Qu\'er\'e,~D.; Wang,~Z.
  Tip-induced flipping of droplets on Janus pillars: From local reconfiguration
  to global transport. \emph{Sci. Adv.} \textbf{2020}, \emph{6}, eabb5440\relax
\mciteBstWouldAddEndPuncttrue
\mciteSetBstMidEndSepPunct{\mcitedefaultmidpunct}
{\mcitedefaultendpunct}{\mcitedefaultseppunct}\relax
\EndOfBibitem
\bibitem[Feng \latin{et~al.}(2021)Feng, Zhu, Zheng, Zhan, Chen, Li, Wang, Yao,
  Liu, and Wang]{Feng2021}
Feng,~S.; Zhu,~P.; Zheng,~H.; Zhan,~H.; Chen,~C.; Li,~J.; Wang,~L.; Yao,~X.;
  Liu,~Y.; Wang,~Z. Three dimensional capillary ratchet-induced liquid
  directional steering. \emph{Science} \textbf{2021}, \emph{373},
  1344--1348\relax
\mciteBstWouldAddEndPuncttrue
\mciteSetBstMidEndSepPunct{\mcitedefaultmidpunct}
{\mcitedefaultendpunct}{\mcitedefaultseppunct}\relax
\EndOfBibitem
\bibitem[Sun \latin{et~al.}(2019)Sun, Wang, Li, Zhang, Ye, Cui, Chen, Wang,
  Butt, Vollmer, and Deng]{Sun2019}
Sun,~Q.; Wang,~D.; Li,~Y.; Zhang,~J.; Ye,~S.; Cui,~J.; Chen,~L.; Wang,~Z.;
  Butt,~H.~J.; Vollmer,~D.; Deng,~X. Surface charge printing for programmed
  droplet transport. \emph{Nat. Mater.} \textbf{2019}, \emph{18},
  936--941\relax
\mciteBstWouldAddEndPuncttrue
\mciteSetBstMidEndSepPunct{\mcitedefaultmidpunct}
{\mcitedefaultendpunct}{\mcitedefaultseppunct}\relax
\EndOfBibitem
\bibitem[Jin \latin{et~al.}(2022)Jin, Xu, Zhang, Li, Sun, Yang, Liu, Mao, and
  Wang]{Jin2022}
Jin,~Y.; Xu,~W.; Zhang,~H.; Li,~R.; Sun,~J.; Yang,~S.; Liu,~M.; Mao,~H.;
  Wang,~Z. Electrostatic tweezer for droplet manipulation. \emph{Proc. Natl.
  Acad. Sci. U.S.A.} \textbf{2022}, \emph{119}, e2105459119\relax
\mciteBstWouldAddEndPuncttrue
\mciteSetBstMidEndSepPunct{\mcitedefaultmidpunct}
{\mcitedefaultendpunct}{\mcitedefaultseppunct}\relax
\EndOfBibitem
\bibitem[Xu \latin{et~al.}(2022)Xu, Jin, Li, Song, Gao, Zhang, Wang, Cui, Yan,
  and Wang]{Xu2022}
Xu,~W.; Jin,~Y.; Li,~W.; Song,~Y.; Gao,~S.; Zhang,~B.; Wang,~L.; Cui,~M.;
  Yan,~X.; Wang,~Z. Triboelectric wetting for continuous droplet transport.
  \emph{Sci. Adv.} \textbf{2022}, \emph{8}, eade2085\relax
\mciteBstWouldAddEndPuncttrue
\mciteSetBstMidEndSepPunct{\mcitedefaultmidpunct}
{\mcitedefaultendpunct}{\mcitedefaultseppunct}\relax
\EndOfBibitem
\bibitem[Zhang \latin{et~al.}(2022)Zhang, Li, Fang, Lin, Zhao, Li, Liu, Chen,
  Lv, and Feng]{Zhang2022}
Zhang,~K.; Li,~J.; Fang,~W.; Lin,~C.; Zhao,~J.; Li,~Z.; Liu,~Y.; Chen,~S.;
  Lv,~C.; Feng,~X.-Q. An energy-conservative many-body dissipative particle
  dynamics model for thermocapillary drop motion. \emph{Phys. Fluids}
  \textbf{2022}, \emph{34}, 052011\relax
\mciteBstWouldAddEndPuncttrue
\mciteSetBstMidEndSepPunct{\mcitedefaultmidpunct}
{\mcitedefaultendpunct}{\mcitedefaultseppunct}\relax
\EndOfBibitem
\bibitem[Dundas \latin{et~al.}(2009)Dundas, McEniry, and Todorov]{Dundas2009}
Dundas,~D.; McEniry,~E.~J.; Todorov,~T.~N. Current-driven atomic waterwheels.
  \emph{Nat. Nanotechnol.} \textbf{2009}, \emph{4}, 99--102\relax
\mciteBstWouldAddEndPuncttrue
\mciteSetBstMidEndSepPunct{\mcitedefaultmidpunct}
{\mcitedefaultendpunct}{\mcitedefaultseppunct}\relax
\EndOfBibitem
\bibitem[Regan \latin{et~al.}(2004)Regan, Aloni, Ritchie, Dahmen, and
  Zettl]{Regan2004}
Regan,~B.~C.; Aloni,~S.; Ritchie,~R.~O.; Dahmen,~U.; Zettl,~A. Carbon nanotubes
  as nanoscale mass conveyors. \emph{Nature} \textbf{2004}, \emph{428},
  924\relax
\mciteBstWouldAddEndPuncttrue
\mciteSetBstMidEndSepPunct{\mcitedefaultmidpunct}
{\mcitedefaultendpunct}{\mcitedefaultseppunct}\relax
\EndOfBibitem
\bibitem[Zhao \latin{et~al.}(2010)Zhao, Huang, Wei, and Zhu]{Zhao2010}
Zhao,~J.; Huang,~J.-Q.; Wei,~F.; Zhu,~J. Mass transportation mechanism in
  electric-biased carbon nanotubes. \emph{Nano Lett.} \textbf{2010}, \emph{10},
  4309--4315\relax
\mciteBstWouldAddEndPuncttrue
\mciteSetBstMidEndSepPunct{\mcitedefaultmidpunct}
{\mcitedefaultendpunct}{\mcitedefaultseppunct}\relax
\EndOfBibitem
\bibitem[Kudernac \latin{et~al.}(2011)Kudernac, Ruangsupapichat, Parschau,
  Maci\'a, Katsonis, Harutyunyan, Ernst, and Feringa]{Kudernac2011}
Kudernac,~T.; Ruangsupapichat,~N.; Parschau,~M.; Maci\'a,~B.; Katsonis,~N.;
  Harutyunyan,~S.~R.; Ernst,~K.-H.; Feringa,~B.~L. Electrically driven
  directional motion of a four-wheeled molecule on a metal surface.
  \emph{Nature} \textbf{2011}, \emph{479}, 208--211\relax
\mciteBstWouldAddEndPuncttrue
\mciteSetBstMidEndSepPunct{\mcitedefaultmidpunct}
{\mcitedefaultendpunct}{\mcitedefaultseppunct}\relax
\EndOfBibitem
\bibitem[Shklyaev \latin{et~al.}(2013)Shklyaev, Mockensturm, and
  Crespi]{Shklyaev2013}
Shklyaev,~O.~E.; Mockensturm,~E.; Crespi,~V.~H. Theory of carbomorph cycles.
  \emph{Phys. Rev. Lett.} \textbf{2013}, \emph{110}, 156803\relax
\mciteBstWouldAddEndPuncttrue
\mciteSetBstMidEndSepPunct{\mcitedefaultmidpunct}
{\mcitedefaultendpunct}{\mcitedefaultseppunct}\relax
\EndOfBibitem
\bibitem[Fennimore \latin{et~al.}(2003)Fennimore, Yuzvinsky, Han, Fuhrer,
  Cumings, and Zettl]{Fennimore2003}
Fennimore,~A.~M.; Yuzvinsky,~T.~D.; Han,~W.-Q.; Fuhrer,~M.~S.; Cumings,~J.;
  Zettl,~A. Rotational actuators based on carbon nanotubes. \emph{Nature}
  \textbf{2003}, \emph{424}, 408--410\relax
\mciteBstWouldAddEndPuncttrue
\mciteSetBstMidEndSepPunct{\mcitedefaultmidpunct}
{\mcitedefaultendpunct}{\mcitedefaultseppunct}\relax
\EndOfBibitem
\bibitem[Bailey \latin{et~al.}(2008)Bailey, Amanatidis, and
  Lambert]{Bailey2008}
Bailey,~S. W.~D.; Amanatidis,~I.; Lambert,~C.~J. Carbon nanotube electron
  windmills: A novel design for nanomotors. \emph{Phys. Rev. Lett.}
  \textbf{2008}, \emph{100}, 256802\relax
\mciteBstWouldAddEndPuncttrue
\mciteSetBstMidEndSepPunct{\mcitedefaultmidpunct}
{\mcitedefaultendpunct}{\mcitedefaultseppunct}\relax
\EndOfBibitem
\bibitem[Huang \latin{et~al.}(2014)Huang, Zhu, and Li]{Huang2014}
Huang,~Y.; Zhu,~S.; Li,~T. {Directional transport of molecular mass on graphene
  by straining}. \emph{Extreme Mech. Lett.} \textbf{2014}, \emph{1},
  83--89\relax
\mciteBstWouldAddEndPuncttrue
\mciteSetBstMidEndSepPunct{\mcitedefaultmidpunct}
{\mcitedefaultendpunct}{\mcitedefaultseppunct}\relax
\EndOfBibitem
\bibitem[Santos; and Ondarquhus(1995)Santos;, and Ondarquhus]{Santos1995}
Santos;,~F.~D.; Ondarquhus,~T. {Free-Running Droplets}. \emph{Phys. Rev. Lett.}
  \textbf{1995}, \emph{75}, 2972\relax
\mciteBstWouldAddEndPuncttrue
\mciteSetBstMidEndSepPunct{\mcitedefaultmidpunct}
{\mcitedefaultendpunct}{\mcitedefaultseppunct}\relax
\EndOfBibitem
\bibitem[Lee \latin{et~al.}(2002)Lee, Kwok, and Laibinis]{Lee2002}
Lee,~S.~W.; Kwok,~D.~Y.; Laibinis,~P.~E. {Chemical influences on
  adsorption-mediated self-propelled drop movement}. \emph{Phys. Rev. E}
  \textbf{2002}, \emph{65}, 9\relax
\mciteBstWouldAddEndPuncttrue
\mciteSetBstMidEndSepPunct{\mcitedefaultmidpunct}
{\mcitedefaultendpunct}{\mcitedefaultseppunct}\relax
\EndOfBibitem
\bibitem[Daniel and Chaudhury(2002)Daniel, and Chaudhury]{Daniel2002}
Daniel,~S.; Chaudhury,~M.~K. Rectified motion of liquid drops on gradient
  surfaces induced by vibration. \textbf{2002}, \emph{18}, 3404--3407\relax
\mciteBstWouldAddEndPuncttrue
\mciteSetBstMidEndSepPunct{\mcitedefaultmidpunct}
{\mcitedefaultendpunct}{\mcitedefaultseppunct}\relax
\EndOfBibitem
\bibitem[Brunet \latin{et~al.}(2007)Brunet, Eggers, and Deegan]{Brunet2007}
Brunet,~P.; Eggers,~J.; Deegan,~R.~D. {Vibration-induced climbing of drops}.
  \emph{Phys. Rev. Lett.} \textbf{2007}, \emph{99}, 3--6\relax
\mciteBstWouldAddEndPuncttrue
\mciteSetBstMidEndSepPunct{\mcitedefaultmidpunct}
{\mcitedefaultendpunct}{\mcitedefaultseppunct}\relax
\EndOfBibitem
\bibitem[Brunet \latin{et~al.}(2009)Brunet, Eggers, and Deegan]{Brunet2009}
Brunet,~P.; Eggers,~J.; Deegan,~R.~D. {Motion of a drop driven by substrate
  vibrations}. \emph{Eur. Phys. J.: Spec. Top.} \textbf{2009}, \emph{166},
  11--14\relax
\mciteBstWouldAddEndPuncttrue
\mciteSetBstMidEndSepPunct{\mcitedefaultmidpunct}
{\mcitedefaultendpunct}{\mcitedefaultseppunct}\relax
\EndOfBibitem
\bibitem[Kwon \latin{et~al.}(2023)Kwon, Kim, Kim, Kim, and Kang]{Kwon2022}
Kwon,~O.~K.; Kim,~J.~M.; Kim,~H.~W.; Kim,~K.~S.; Kang,~J.~W. A Study on
  Nanosensor Based on Graphene Nanoflake Transport on Graphene Nanoribbon Using
  Edge Vibration. \emph{J. Electr. Eng. Technol.} \textbf{2023}, \emph{18},
  663--668\relax
\mciteBstWouldAddEndPuncttrue
\mciteSetBstMidEndSepPunct{\mcitedefaultmidpunct}
{\mcitedefaultendpunct}{\mcitedefaultseppunct}\relax
\EndOfBibitem
\bibitem[Buguin \latin{et~al.}(2002)Buguin, Talini, and Silberzan]{Buguin2002}
Buguin,~A.; Talini,~L.; Silberzan,~P. {Ratchet-like topological structures for
  the control of microdrops}. \emph{Appl. Phys. A: Mater. Sci. Process.}
  \textbf{2002}, \emph{75}, 207--212\relax
\mciteBstWouldAddEndPuncttrue
\mciteSetBstMidEndSepPunct{\mcitedefaultmidpunct}
{\mcitedefaultendpunct}{\mcitedefaultseppunct}\relax
\EndOfBibitem
\bibitem[Thiele and John(2010)Thiele, and John]{Thiele2010}
Thiele,~U.; John,~K. {Transport of free surface liquid films and drops by
  external ratchets and self-ratcheting mechanisms}. \emph{Chem. Phys.}
  \textbf{2010}, \emph{375}, 578--586\relax
\mciteBstWouldAddEndPuncttrue
\mciteSetBstMidEndSepPunct{\mcitedefaultmidpunct}
{\mcitedefaultendpunct}{\mcitedefaultseppunct}\relax
\EndOfBibitem
\bibitem[Noblin \latin{et~al.}(2009)Noblin, Kofman, and Celestini]{Noblin2009}
Noblin,~X.; Kofman,~R.; Celestini,~F. {Ratchetlike motion of a shaken drop}.
  \emph{Phys. Rev. Lett.} \textbf{2009}, \emph{102}, 1--4\relax
\mciteBstWouldAddEndPuncttrue
\mciteSetBstMidEndSepPunct{\mcitedefaultmidpunct}
{\mcitedefaultendpunct}{\mcitedefaultseppunct}\relax
\EndOfBibitem
\bibitem[Ni \latin{et~al.}(2022)Ni, Song, Li, Lu, Jiang, and Li]{Ni2022}
Ni,~E.; Song,~L.; Li,~Z.; Lu,~G.; Jiang,~Y.; Li,~H. Unidirectional
  self-actuation transport of a liquid metal nanodroplet in a two-plate
  confinement microchannel. \emph{Nanoscale Adv.} \textbf{2022}, \emph{4},
  2752--2761\relax
\mciteBstWouldAddEndPuncttrue
\mciteSetBstMidEndSepPunct{\mcitedefaultmidpunct}
{\mcitedefaultendpunct}{\mcitedefaultseppunct}\relax
\EndOfBibitem
\bibitem[Barthlott \latin{et~al.}(2016)Barthlott, Mail, and
  Neinhuis]{Barthlott2016}
Barthlott,~W.; Mail,~M.; Neinhuis,~C. Superhydrophobic hierarchically
  structured surfaces in biology: evolution, structural principles and
  biomimetic applications. \emph{Phil. Trans. R. Soc. A} \textbf{2016},
  \emph{374}, 2016019\relax
\mciteBstWouldAddEndPuncttrue
\mciteSetBstMidEndSepPunct{\mcitedefaultmidpunct}
{\mcitedefaultendpunct}{\mcitedefaultseppunct}\relax
\EndOfBibitem
\bibitem[Badr \latin{et~al.}(2022)Badr, Hauer, Vollmer, and Schmid]{Badr2022}
Badr,~R. G.~M.; Hauer,~L.; Vollmer,~D.; Schmid,~F. Cloaking Transition of
  Droplets on Lubricated Brushes. \emph{J. Phys. Chem. B} \textbf{2022},
  \emph{126}, 7047--7058\relax
\mciteBstWouldAddEndPuncttrue
\mciteSetBstMidEndSepPunct{\mcitedefaultmidpunct}
{\mcitedefaultendpunct}{\mcitedefaultseppunct}\relax
\EndOfBibitem
\bibitem[Button \latin{et~al.}(2012)Button, Cai, Ehre, Kesimer, Hill, Sheehan,
  Boucher, and Rubinstein]{Button2012}
Button,~B.; Cai,~L.-H.; Ehre,~C.; Kesimer,~M.; Hill,~D.~B.; Sheehan,~J.~K.;
  Boucher,~R.~C.; Rubinstein,~M. A Periciliary Brush Promotes the Lung Health
  by Separating the Mucus Layer from Airway Epithelia. \emph{Science}
  \textbf{2012}, \emph{337}, 937--941\relax
\mciteBstWouldAddEndPuncttrue
\mciteSetBstMidEndSepPunct{\mcitedefaultmidpunct}
{\mcitedefaultendpunct}{\mcitedefaultseppunct}\relax
\EndOfBibitem
\bibitem[Stukowski(2010)]{Stukowski2010}
Stukowski,~A. Visualization and analysis of atomistic simulation data with
  OVITO–the Open Visualization Tool. \emph{Modelling Simul. Mater. Sci. Eng.}
  \textbf{2010}, \emph{18}, 015012\relax
\mciteBstWouldAddEndPuncttrue
\mciteSetBstMidEndSepPunct{\mcitedefaultmidpunct}
{\mcitedefaultendpunct}{\mcitedefaultseppunct}\relax
\EndOfBibitem
\bibitem[Kremer and Grest(1990)Kremer, and Grest]{Kremer1990}
Kremer,~K.; Grest,~G.~S. Dynamics of entangled linear polymer melts: A
  molecular-dynamics simulation. \emph{J. Chem. Phys.} \textbf{1990},
  \emph{92}, 5057\relax
\mciteBstWouldAddEndPuncttrue
\mciteSetBstMidEndSepPunct{\mcitedefaultmidpunct}
{\mcitedefaultendpunct}{\mcitedefaultseppunct}\relax
\EndOfBibitem
\bibitem[Theodorakis and Fytas(2011)Theodorakis, and Fytas]{Theodorakis2011}
Theodorakis,~P.~E.; Fytas,~N.~G. Microphase separation in linear multiblock
  copolymers under poor solvent conditions. \emph{Soft Matter} \textbf{2011},
  \emph{7}, 1038--1044\relax
\mciteBstWouldAddEndPuncttrue
\mciteSetBstMidEndSepPunct{\mcitedefaultmidpunct}
{\mcitedefaultendpunct}{\mcitedefaultseppunct}\relax
\EndOfBibitem
\bibitem[Tretyakov and M{\"{u}}ller(2014)Tretyakov, and
  M{\"{u}}ller]{Tretyakov2014}
Tretyakov,~N.; M{\"{u}}ller,~M. {Directed transport of polymer drops on
  vibrating superhydrophobic substrates: a molecular dynamics study.}
  \emph{Soft matter} \textbf{2014}, \emph{10}, 4373--86\relax
\mciteBstWouldAddEndPuncttrue
\mciteSetBstMidEndSepPunct{\mcitedefaultmidpunct}
{\mcitedefaultendpunct}{\mcitedefaultseppunct}\relax
\EndOfBibitem
\bibitem[Theodorakis \latin{et~al.}(2010)Theodorakis, Paul, and
  Binder]{Theodorakis2010a}
Theodorakis,~P.~E.; Paul,~W.; Binder,~K. Pearl-necklace structures of molecular
  brushes with rigid backbone under poor solvent conditions: A simulation
  study. \emph{J. Chem. Phys.} \textbf{2010}, \emph{133}, 104901\relax
\mciteBstWouldAddEndPuncttrue
\mciteSetBstMidEndSepPunct{\mcitedefaultmidpunct}
{\mcitedefaultendpunct}{\mcitedefaultseppunct}\relax
\EndOfBibitem
\bibitem[Theodorakis \latin{et~al.}(2010)Theodorakis, Paul, and
  Binder]{Theodorakis2010b}
Theodorakis,~P.~E.; Paul,~W.; Binder,~K. Interplay between Chain Collapse and
  Microphase Separation in Bottle-Brush Polymers with Two Types of Side Chains.
  \emph{Macromolecules} \textbf{2010}, \emph{43}, 5137--5148\relax
\mciteBstWouldAddEndPuncttrue
\mciteSetBstMidEndSepPunct{\mcitedefaultmidpunct}
{\mcitedefaultendpunct}{\mcitedefaultseppunct}\relax
\EndOfBibitem
\bibitem[Schneider and Stoll(1978)Schneider, and Stoll]{Schneider1978}
Schneider,~T.; Stoll,~E. Molecular-dynamics study of a three-dimensional
  one-component model for distortive phase transitions. \emph{Phys. Rev. B}
  \textbf{1978}, \emph{17}, 1302--1322\relax
\mciteBstWouldAddEndPuncttrue
\mciteSetBstMidEndSepPunct{\mcitedefaultmidpunct}
{\mcitedefaultendpunct}{\mcitedefaultseppunct}\relax
\EndOfBibitem
\bibitem[Plimpton(1995)]{Plimpton1995}
Plimpton,~S. Fast Parallel Algorithms for Short-Range Molecular Dynamics.
  \emph{J. Comp. Phys.} \textbf{1995}, \emph{117}, 1--19\relax
\mciteBstWouldAddEndPuncttrue
\mciteSetBstMidEndSepPunct{\mcitedefaultmidpunct}
{\mcitedefaultendpunct}{\mcitedefaultseppunct}\relax
\EndOfBibitem
\bibitem[Barber \latin{et~al.}(1996)Barber, Dobkin, and Huhdanpaa]{qhull1996}
Barber,~C.~B.; Dobkin,~D.~P.; Huhdanpaa,~H.~T. The Quickhull algorithm for
  convex hulls. \emph{ACM Trans. on Mathematical Software} \textbf{1996},
  \emph{22}, 469--483, http://www.qhull.org\relax
\mciteBstWouldAddEndPuncttrue
\mciteSetBstMidEndSepPunct{\mcitedefaultmidpunct}
{\mcitedefaultendpunct}{\mcitedefaultseppunct}\relax
\EndOfBibitem
\bibitem[Theodorakis \latin{et~al.}(2015)Theodorakis, M{\"u}ller, Craster, and
  Matar]{Theodorakis2015}
Theodorakis,~P.~E.; M{\"u}ller,~E.~A.; Craster,~R.~V.; Matar,~O.~K. Modelling
  the superspreading of surfactant-laden droplets with computer simulation.
  \emph{Soft Matter} \textbf{2015}, \emph{11}, 9254--9261\relax
\mciteBstWouldAddEndPuncttrue
\mciteSetBstMidEndSepPunct{\mcitedefaultmidpunct}
{\mcitedefaultendpunct}{\mcitedefaultseppunct}\relax
\EndOfBibitem
\end{mcitethebibliography}
\end{document}